\documentclass[sigconf]{acmart}
\renewcommand\footnotetextcopyrightpermission[1]{} 

\renewcommand\footnotetextcopyrightpermission[1]{} 
\RequirePackage{etex}
\AtBeginDocument{%
  \providecommand\BibTeX{{%
    \normalfont B\kern-0.5em{\scshape i\kern-0.25em b}\kern-0.8em\TeX}}}

\usepackage{amsmath}
\usepackage{algorithmic}
\usepackage{graphicx}
\usepackage{textcomp}
\usepackage{xcolor}
\usepackage{multicol}
\usepackage{framed}
\usepackage{booktabs}
\usepackage{tabularx}
\usepackage{mathtools}
\usepackage{multirow}
\usepackage{pgfplots}
\usepackage{tikz}
\usepackage{float}
\usepackage{appendix}
\usepackage{enumitem}

\usepackage[misc]{ifsym}

\usepackage{pifont}
\newcommand{\cmark}{\ding{51}}%
\newcommand{\xmark}{\ding{55}}%

\usepackage{balance}
\usepackage{url}
\usetikzlibrary{patterns}
\pgfplotsset{compat=newest}
\usetikzlibrary{plotmarks}
\usetikzlibrary{arrows.meta}
\usepgfplotslibrary{patchplots}
\usepackage{grffile}
\usetikzlibrary{patterns}
\usepackage{balance}
\usepackage{bookmark}
\usepackage[
n,
advantage,
operators,
sets,
adversary,
landau,
probability,
notions,
logic,
ff,
mm,
primitives,
events,
complexity,
asymptotics,
keys
]{cryptocode}

\usepackage[utf8]{inputenc}
\usepackage[english]{babel}
\usepackage{amsthm}

\newtheorem{theorem}{\textbf{Theorem}}
\newtheorem{definition}{\textbf{Definition}}
\usepackage[normalem]{ulem}

\newcommand{\idealMain}{\texorpdfstring{$\mathcal{F}_{\TheName{}}$}{Lg}}

\newcommand{\idealGatt}{\texorpdfstring{$\mathcal{G}_{att}$}}
\newcommand{\idealBC}{\texorpdfstring{$\mathcal{F}_{blockchain}$}}
\newcommand{\protSpeedster}{\texorpdfstring{$\Pi_{\TheName{}}$}{Lg}}
\newcommand{\TheName}{\textsc{Speedster}}

\newcommand{\aveChannel}{\text{$c$}}
\newcommand{\spstate}{\sffont{state}}

\newcommand{\ASE}{\texorpdfstring{$\mathcal{AE}$}}

\newcommand{\certChannel}{\textit{Certified Channel}}

\newcommand{\msk}{\text{\footnotesize{\textsf{msk}}}}
\newcommand{\mpk}{\text{\footnotesize{\textsf{mpk}}}}

\newcommand{\sffont}[1]{\text{\footnotesize{\textsf{#1}}}}
\newcommand{\mathcolor}[1]{\textcolor{black}{#1}}

\newcommand{\colorgray}[1]{\textit{\textcolor{black}{#1}}}

\newcommand{\fund}{\text{\footnotesize{\textsf{fund}}}}
\newcommand{\acc}{\text{\footnotesize{\textsf{acc}$_{enclave}$}}}
\newcommand{\cert}{\text{\footnotesize{\textsf{cert}}}}
\newcommand{\ccid}{\text{\footnotesize{\textsf{ccid}}}}
\newcommand{\cid}{\text{\footnotesize{\textsf{cid}}}}
\newcommand{\tx}{\text{\footnotesize{\textsf{tx}}}}

\newcommand{\msg}{\text{\footnotesize{\textsf{aux}}}}
\newcommand{\inp}{\text{\footnotesize{\textsf{inp}}}}
\newcommand{\outp}{\text{\footnotesize{\textsf{outp}}}}
\newcommand{\ct}{\text{\footnotesize{\textsf{ct}}}}
\newcommand{\ck}{\text{\footnotesize{\textsf{ck}}}}
\newcommand{\cp}{\text{\footnotesize{\textsf{cp}}}}

\newcommand{\prog}{$\sffont{prog}_{enclave}$}

\newcommand{\contract}{\sffont{contract}$_{\TheName{}}$}

\newcommand{\idealE}{\text{\textcolor{black}{$\mathcal{E}$}}}
\newcommand{\idealP}{\text{\textcolor{black}{$\mathcal{P}$}}}
\newcommand{\idealPi}{\text{\textcolor{black}{$\mathcal{P}_i$}}}
\newcommand{\idealPj}{\text{\textcolor{black}{$\mathcal{P}_j$}}}
\newcommand{\idealR}{\text{\textcolor{black}{$\mathcal{R}$}}}
\newcommand{\idealS}{\text{\textcolor{black}{$\mathcal{S}$}}}
\newcommand{\idealA}{\text{\textcolor{black}{\adv{}}}}

\newcommand{\GCM}{\text{\texttt{AES-GCM}}}
\newcommand{\ECDSA}{\text{\texttt{ECDSA}}}

\begin{document}

\title{\TheName{}: A TEE-assisted State Channel System}

\thanks{\textsuperscript{\Letter}~
{Fengwei Zhang is the corresponding author.}}

\author{Jinghui Liao\textsuperscript{1}, Fengwei Zhang\textsuperscript{2,  3,  \Letter}, Wenhai Sun\textsuperscript{4}, Weisong Shi\textsuperscript{1}}

\affiliation{
 \institution{
   \textsuperscript{1}\textit{Department of Computer Science,  Wayne State University} \\
 \textsuperscript{2}\textit{Department of Computer Science and Engineering,  Southern University of Science and Technology} \\
\textsuperscript{3}\textit{Research Institute of Trustworthy Autonomous Systems,  Southern University of Science and Technology} \\
 \textsuperscript{4}\textit{Department of Computer and Information Technology,  Purdue University} \\
jinghui@wayne.edu,  zhangfw@sustech.edu.cn,  whsun@purdue.edu,  weisong@wayne.edu \\
 }
 \country{}
}

\keywords{Multi-party State Channel, Layer-2, Scalability, Offchain Smart Contracts}

\begin{abstract}
State channel network is the most popular layer-2 solution to the issues of scalability, high transaction fees, and low transaction throughput of public Blockchain networks. However, the existing works have limitations that curb the wide adoption of the technology, such as the expensive creation and closure of channels, strict synchronization between the main chain and off-chain channels, frozen deposits, and inability to execute multi-party smart contracts. 

In this work, we present \textsc{Speedster}, an account-based state-channel system that aims to address the above issues. To this end, \textsc{Speedster} leverages the latest development of secure hardware to create dispute-free \textit{certified channels} that can be operated efficiently off the Blockchain. 
\textsc{Speedster} is fully decentralized and provides better privacy protection. It supports fast native multi-party contract execution, which is missing in prior TEE-enabled channel networks. 
Compared to the Lightning Network, \textsc{Speedster} improves the throughput by about $10,000\times$ and generates $97\%$ less on-chain data with a comparable network scale.

\end{abstract}

\maketitle

\pagestyle{plain}

\section{Introduction}
\label{sec:introduction}
Blockchain (\textit{aka} layer-1 main chain) has been deemed a disruptive technology to build decentralized trust and foster innovative applications in both public and private sectors. However, scalability has become a great concern in practice when adopting the decentralized infrastructure. For example, the Bitcoin network \cite{nakamoto2008bitcoin} can only handle approximately $3,500$ transactions in every new block due to the block size limitation \cite{btc2019size} and process 7 transactions per second ($tps$) on average~\cite{nakamoto2008bitcoin,btc2019tps}. The issue has also haunted other major Blockchain networks which are based on a similar design principle, such as Ethereum~\cite{buterin2014ethereum}. Modifying the on-chain protocols helps alleviate the problem, for instance, using alternative consensus algorithms \cite{mingxiao2017review} and improving the information propagation \cite{zamani2018rapidchain,luu2016secure}.  
Nevertheless, changes at layer-1 Blockchain level may adversely affect the existing participants with undesired costs \cite{bch2019Hardfork,eth2019upgrade}. 
Shifting to layer-2 payment channels~\cite{lightNn, Raiden, miller2019sprites, burchert2018scalable} is considered an effective remedy by carrying out micropayment transactions off the Blockchain to avoid the expensive on-chain overhead. State channels \cite{miller2019sprites, allison2016ethereum, dziembowski2018general} further advances this off-chain innovation by enabling stateful transactions and smart contract execution. Promising as it is, the state channel also has the following limitations.

\vspace{3pt}
{(\textbf{L1})} Opening a new channel needs to freeze the deposit of channel participants during the whole life cycle of the channel, which significantly affects the liquidity and network effectiveness. Every time a channel is created or closed, an associated transaction is required to signal the main chain, thus incurring additional transaction fees and waiting time for block confirmation \cite{lightNn, Raiden, miller2019sprites, lind2016teechan, allison2016ethereum}.

\vspace{3pt}
{(\textbf{L2})} The current dispute resolution in the state channel is not robust and vulnerable to the denial-of-service (DoS) attack. A malicious channel participant can send an outdated channel state to the Blockchain while DoS-ing the victim to prevent the submission of the lasted channel state.

\vspace{3pt}
{(\textbf{L3})} With the help of Hashed Timelock Contract (HTLC)~\cite{lightNn}, the architectural complexity is reduced and multi-hop transaction becomes feasible in the state channel network. However, HTLC also raises many privacy concerns with the intermediate nodes~\cite{green2017bolt, malavolta2017concurrency, dziembowski2019perun, herrera2019difficulty} and leads to a multitude of attacks, such as wormhole attacks~\cite{malavolta2019anonymous}, bribery  attacks~\cite{tsabary2020mad}, and DoS attacks~\cite{tochner2019hijacking, khalil2017revive}.

\vspace{3pt}
{(\textbf{L4})} Despite the ambition of the instant processing of off-chain transactions~\cite{lightNn}, the complex routing and state updating mechanisms give rise to a non-negligible overhead, thus considerably degrading promised performance. The actual throughput of the state channels is still unsatisfactory (tens of $tps$ measured in  \cite{miller2019sprites, lind2019teechain, dziembowski2019perun}).

\vspace{3pt}
{(\textbf{L5})} The state exchange is confined within a pairwise channel, which poses fundamental challenges for creating and executing multi-party smart contracts. Though a multi-party state channel can be recursively established using the virtual channel techniques~\cite{dziembowski2019perun,dziembowski2019multi,close2019nitro}, the associated expensive cost is still a concern for implementation.

\vspace{3pt}
\textbf{\textsc{Technical contributions.}}
We present \TheName{} to address the above limitations. The main idea of \TheName{} is that every user creates and funds an off-chain account protected by the enclave, an instance of a Trusted Execution Environment (TEE). As \TheName{} transfers the on-chain trust with the Blockchain to the off-chain trust with enclaves, we significantly reduce the design complexity to accomplish a plethora of innovations, such as multi-party channels, and lightweight protocols for channel confidentiality, authenticity, finalization, and dispute resolution. \TheName{} outperforms the conventional state channel networks in terms of security, performance, and functionality.

In \TheName{}, a node does not need to send on-chain transaction to open/close a channel.
Only one deposit transaction is needed to initialize a TEE-enabled account for each off-chain participant. Later, a participating node can directly create/close channels with any other nodes completely off the main chain with the balance in their enclave accounts. Therefore, \TheName{} is a fully decentralized state channel network that enjoys instant channel creation/closure and enhanced privacy by eliminating HTLC-based multi-hopping and routing~\cite{malavolta2019anonymous, tochner2019hijacking, khalil2017revive, tsabary2020mad}.

\TheName{} adopts a novel certificate-based off-chain transaction processing model where the channel state is retained in the enclave. \TheName{} modifies the state before sending out or after receiving transactions to make sure the state submitted to the Blockchain is always up to date. As a result, an attacker cannot roll back to old states by DoS-ing counterparts and fool the Blockchain into a biased decision.

Off-chain multi-party smart contracts can be enabled and efficiently executed in \TheName{}. With the certificate-based channels, \TheName{} naturally supports interactions among multiple parties. The state information can be correctly exchanged across multiple channels of the same account.


By leveraging the off-chain enclave trust, \TheName{} replaces the costly public-key algorithms with efficient symmetric-key operations for transaction generation and verification. Experimental results show that \TheName{} increases the throughput by four orders of magnitude compared to the Lightning Network, the most popular payment channel network in practice.


\vspace{7pt}
\textbf{\textsc{Evaluation}.} 

\TheName{} is intentionally designed to be compatible with different major TEE platforms for availability and usability, such as AMD~\cite{AMDSEV}, Intel~\cite{sgx:white1}, and ARM~\cite{ARMTrustZone}. We evaluate its cross-platform performance to show the advantage over other popular layer-2 designs. Specifically, we migrate eEVM~\cite{microsoft2019eEVM}, a full version of Ethereum Virtual Machine (EVM)~\cite{evm} into \TheName{}, and execute unmodified Ethereum smart contracts off-chain. We develop a set of benchmark contracts to show the unique features and performance of \TheName{}. Through thorough experiments, we present the specification for running \TheName{}, the much-improved transaction throughput, and the capability of executing different kinds of smart contracts that traditional state channels cannot support. The experiments include:

\begin{itemize}
\item Transaction load test: To test the transaction throughput directly between two parties without loading any smart contract;
\item Instant state sharing: Participants can update and share their states instantly; this is an important performance indicator for time-sensitive applications, such as racing games and decentralized financial services; 
\item ERC20 contracts: To show the performance of off-chain fund exchange;
\item Gomoku contract: To show the performance of the turn-based contracts;
\item Paper-Scissors-Rock contract: To illustrate the fairness (for in-parallel execution) in \TheName{} channel;
\item Monopoly contract: To test the multi-party state channel capability of \TheName{}, we load a Monopoly smart contract that is executed by four players alternately.
\end{itemize}

The evaluation results show that \TheName{} is efficient and takes only $0.02 ms$, $0.14 ms$, and $20.49 ms$ to process a value-transfer transaction on Intel, AMD, and ARM platforms, respectively, which leads to much higher throughput than that of Lightning network. The source code of \TheName{} is available at \url{https://bit.ly/3a32ju7}. 
\section{Background}

In this section, we provide the background of the technologies that we use in our system.

\subsection{Blockchain and Smart Contract}

\textit{Blockchain} is a distributed ledger that leverages cryptography to maintain a transparent, immutable, and verifiable transaction record ~\cite{nakamoto2008bitcoin, buterin2014ethereum}. In contrast to the permissioned Blockchain~\cite{androulaki2018hyperledger, swanson2015consensus}, permissionless Blockchain~\cite{wust2018you} is publicly accessible but constrained by the inefficient consensus protocols, such as the Nakamoto consensus in Bitcoin \cite{nakamoto2008bitcoin, jakobsson1999proofs}, on top of the asynchronous network infrastructure, which leads to a series of performance bottlenecks in practice. See \cite{urquhart2016inefficiency,gervais2016security,saleh2020blockchain,zhang2017rem} for detailed discussion.

\vspace{2pt}
\textit{Smart contracts} in Blockchain complement the ledger functions by providing essential computations. In general, a smart contract is a program that is stored as a transaction on the Blockchain. Once being called, the contract will be executed by all the nodes in the network. The whole network will verify the computation result through consensus protocols, thus creating a fair and trustless environment to foster a range of novel decentralized applications~\cite{mccorry2017smart,zhang2018smart}. A well-known example is the Ethereum smart contract~\cite{buterin2014ethereum,buterin2014next}, which runs inside the EVM~\cite{evm}. EVM needs to be set up on every full Ethereum node to create an isolated environment from the network, file system, and I/O services for contract execution. The user transactions will be taken as input to the contract inside the EVM.

\subsection{Layer-2 Channels}
\label{subsec:layer2}

Layer-2 technologies are proposed to address the scalability concerns ~\cite{ave2019confirm}, short storage for historical transactions~\cite{blockchainSize}, etc., for the layer-1 Blockchain.  

\vspace{2pt}
\textit{Payment channel} is the first attempt to use an off-chain infrastructure to process micropayments between two parties without frequent main chain involvement. To create a channel, each party needs to send a transaction to the Blockchain to lock in a certain amount of deposit on the main chain until a transaction is issued later to close the channel. When the channel is open, transactions can be sent back and forth between participants as long as they do not surpass the committed channel capacity. 

\vspace{2pt}
\textit{Payment channel network} (PCN) is built on top of the individual payment channels to route transactions for any pair of parties who may not have direct channel connections~\cite{miller2019sprites,khalil2017revive,lightNn}. 
Hashed Timelock Contract is exploited to guarantee balance security along the payment route, i.e., the balances of the involved nodes are changed in compliance with the prescribed agreement. 
PCN greatly relieves the users from costly channel creation and management, but it also brings up concerns about the privacy with intermediate routing nodes and the formation of the centrality of the network.  

\vspace{2pt}
\textit{State channel network} extends PCN by allowing for stateful activities, such as off-chain smart contract~\cite{dziembowski2019perun,dziembowski2018foundations,dziembowski2018general,close2019nitro,dziembowski2019multi}. 
However, recording and updating states across multiple parties are still expensive due to the sophisticated trust management and protocol design. For example, the current multi-party state channel~\cite{close2019nitro,dziembowski2019multi} is realized through recursive virtual channel establishment~\cite{dziembowski2019perun,dziembowski2018foundations,dziembowski2018general}, which introduces non-negligible complexity and overhead.  

\vspace{2pt}
Regardless of the technical differences of the above layer-2 technologies, they all need to involve the inefficient Blockchain for channel creation, closure, or dispute resolution. Moreover, privacy and instability~\cite{tochner2019hijacking} concerns also arise and hamper the wide adoption of those technologies.

\subsection{Trusted Execution Environment} 

\textit{Trusted Execution Environment} provides a secure, isolated environment (or enclave) in a computer system to execute programs with sensitive data. 
Enclave protects the data and code inside against inference and manipulation by other programs outside the trusted computing base (TCB). Intel Software Guard eXtensions (SGX)~\cite{sgx:white1, sgx:white2, sgx:white3} and AMD Secure Encrypted Virtualization (SEV)~\cite{kaplan2016amd, amdsecure} are two popular general-purpose hardware-assisted TEEs developed for the x86 architecture. 
Precisely, the TCB of SGX is a set of new processor instructions and data structures that are introduced to support the execution of the enclave. 
The TCB of AMD SEV is the SEV-enabled virtual machine protected by an embedded 32-bit microcontroller (ARM Cortex-A5)~\cite{kaplan2016amd}. 
Other prominent TEE examples include TrustZone~\cite{ARMTrustZone} and CCA~\cite{armv9} on ARM, MultiZone~\cite{hex2019smbedtls} and KeyStone~\cite{lee2019keystone} on RISC-V, and Apple Secure Enclave in T2 chip~\cite{apple2019T2}.
To demonstrate the cross-platform capability of \TheName{}, we implement a prototype that can run on Intel, AMD, and ARM machines, and we make \TheName{} design general enough for other TEE platforms not limited to the tested environments. 

\vspace{2pt}
\textit{Remote attestation}~\cite{pass2017formal} is used to verify the authenticity of the enclave before executing enclave programs. Specifically, to prevent attackers from simulating the enclave, a TEE-enabled processor uses a hard-coded root key to cryptographically sign the measurement of the enclave, including the initial state, code, and data.
Note that even if one TEE processor sets up multiple enclaves with the same set of functions, their respective measurements will be distinctively different. As such, everyone can publicly verify the authenticity of the established enclave with help from vendors.
\section{Threat Model and Design Goals}
In this section, we present the threat model and design goals of \TheName{}.

\subsection{Threat Model}
\label{subsec:threatmodel} 

We assume that nodes in the system run on TEE-enabled platforms, and all parties trust the enclaves after the successful attestation. 
An adversary may compromise the operating system of a target node and further control the system's software stack. 

In \TheName{}, we use TEE as a secure abstraction to make the design and security independent of the specific platforms. We provide rigorous security proofs to show the reliability and robustness of \TheName{}. 
However, like any secure function, theoretical security could be compromised by erroneous implementations. Therefore, to be consistent with prior work~\cite{lind2019teechain, cheng2018ekiden, das2019fastkitten}, we additionally consider attacks on specific TEE platforms in our implementation for completeness, which does not indicate the insecurity of the general design of \TheName{}. See Section~\ref{subsec:teeattack} for the detailed discussion for the particular platforms.

Similar to prior research~\cite{dziembowski2019perun, dziembowski2018foundations, dziembowski2018general, close2019nitro, dziembowski2019multi}, this work also assumes a Blockchain abstraction to provide desired ledger functions, such as transparent and immutable storage, and verifiable computations with smart contracts. \TheName{} assumes that the Blockchain nodes are equipped with adequate resources for computation and storage so that we only concentrate on the off-chain related design (see Section~\ref{abstraction} for more discussion on the TEE and Blockchain abstractions).

\subsection{Design Goals}
\label{sec:techchallenges}

In this subsection, we elaborate on the main design goals of \TheName{}. 

\vspace{3pt}
\noindent \textbf{Efficient Channel System {(L1, L2, L4)} :}
The current layer-2 channel system design principle derails from the promised efficiency for off-chain micropayment processing. 
As discussed in Section~\ref{subsec:layer2}, the existing systems need expensive interactions with the Blockchain for various channel operations in terms of time and economic costs. Users are required to trust the intermediate nodes and pay extra fees for transaction forwarding and state updating. 

In this work, we attempt to devise a functionally efficient off-chain network that aims to significantly reduce the channel cost for creation and closure and eliminate the dispute in light of unsynchronized communications.

\vspace{3pt}
\noindent \textbf{Fully Decentralized Channel Network {(L3, L4)}:}
Due to the expensive channel cost, a node in layer-2 currently cannot afford to establish direct channel connections with all other nodes in the system. Multi-hopping addresses the problem but raises privacy concerns about the emergent centralized payment hubs~\cite{herrera2019difficulty,rohrer2019discharged,dziembowski2019perun}, which is at odds with the decentralization promise of Blockchain. 

In contrast, we aim to build a fully decentralized channel network to allow users to freely set up direct channels with intended parties, thus eliminating centrality concerns. Note that none of the existing work can support this function ~\cite{lightNn,miller2019sprites,lind2016teechan,lind2019teechain}.
 
\vspace{3pt}
\noindent \textbf{Efficient Multi-Party State Channel {(L5)}:}
\label{subsec:multiparty}
Sharing states among multiple parties is instrumental for many real-world applications, such as voting, auctioning, and gaming. However, most off-chain state channels only support pairwise state exchange~\cite{dziembowski2019perun, dong2018celer}. The involvement of more channel participants depends on intermediaries, which complicates the network setup and trust management~\cite{dziembowski2019multi, close2019nitro}.

\TheName{} targets a more efficient multi-party state channel by streamlining the architectural design for easy setup and use. The state information of one \TheName{} node can be freely shared with other parties of interest without worrying about the additional cost in prior work.

\vspace{3pt}
\noindent \textbf{Other Goals:}
Besides, \TheName{} also aims to: (1) preserve the privacy of transactions (see Section~\ref{subsec:security} for detailed security definition and analysis), (2) be abstract and general enough to not rely on any specific TEE platform. 

\section{\TheName{} Design}
\label{sec:overview}
In this section, we first introduce the architecture of \TheName{} and then detail the design.

\subsection{System Architecture}
\input{paragraphs/figure/structure.tex}

\TheName{} contains two components: the state channel core program \prog{} executed inside the enclave and the on-chain smart contract \contract{} running on the Blockchain. Figure~\ref{fig:arch-ASCS} shows the high-level architecture of \TheName{}, in which two participants are connected by a \certChannel{} (see Definition~\ref{def:def_auth}). 

\vspace{2pt}
\noindent\textbf{Prog$_{enclave}$}. The program runs in the enclave. \prog{} creates and manages an enclave account for a \TheName{} node. It executes commands from the user to open and close channels as well as constructs and processes channel transactions. To verify the enclave authenticity, it also generates measurements for remote attestation.

\vspace{2pt}
\noindent\textbf{Contract$_{\TheName{}}$}. \contract{} is a smart contract deployed on the Blockchain to manage the on-chain state of \TheName{} account. To register an account, a deposit has to be sent to this contract and recorded in the Blockchain. Later, the information can be used to initialize the enclave state. It also handles transactions to claim \fund{} for \TheName{} account.

\subsection{Workflow}

\begin{figure}[htp]
    \centering
    \includegraphics[width=7cm]{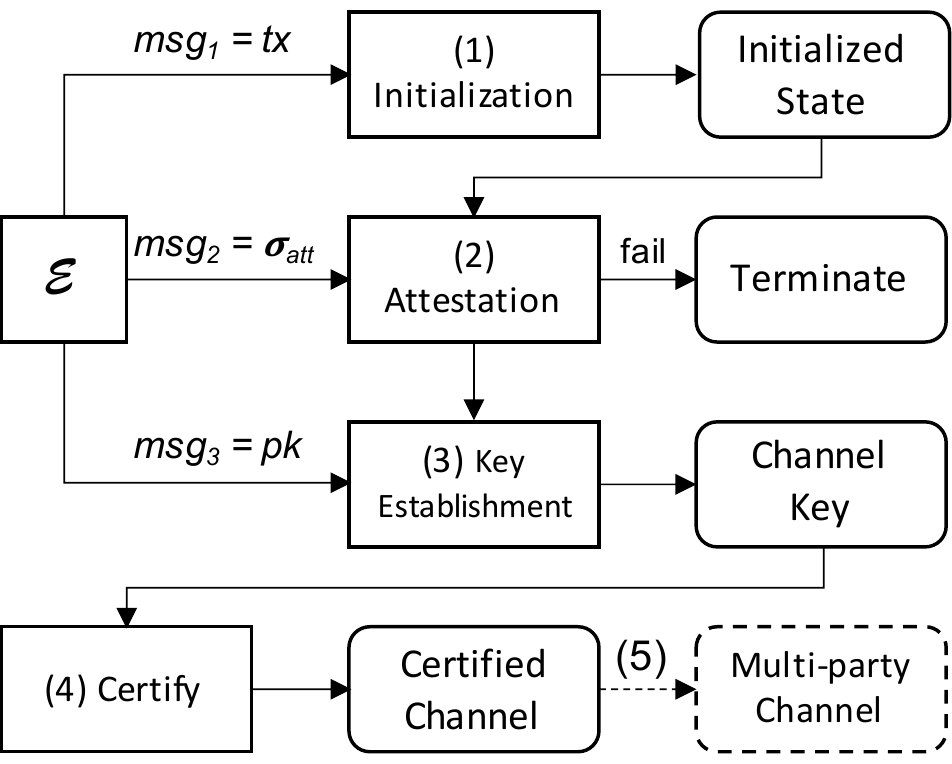}
    \caption{Workflow of node initialization and certified channel creation. $\mathcal{E}$ is the environment, including the Blockchain and the channel users, who pass input to \TheName{} nodes.}
    \label{fig:workflow}
\end{figure}

In this subsection, we outline the workflow of \TheName{} including: (1) node initialization, (2) enclave state attestation, (3) channel key establishment, (4) channel certification, and (5) multi-party state channel establishment (optional). The workflow is illustrated in Figure~\ref{fig:workflow}.

\vspace{3pt}
\textit{Node Initialization:}
In this step, an account \acc{} along with a pair of keys \pk{} and \sk{} is opened in the enclave after the program \prog{} is loaded into the enclave for the first time. The enclave keeps \sk{} private and publishes \pk{} as the account address that can be used to deposit \acc{} on the Blockchain. To ensure the authenticity of the opened account \acc{} for following off-chain attestations, an initial deposit transaction needs to be generated to register \acc{} with the Blockchain. 
After the Blockchain confirms the transaction, the user loads relevant information into the enclave as a registration proof to initialize the enclave state $\spstate^0 := (\tx_0, \msg_0)$, a tuple that contains the deposit transaction $\tx{}_0$ and auxiliary information $\msg_0$, where $\tx{}_0$ can be more than one deposit and $\msg_0$ can be the current balance or account-related configuration information. Further deposits are allowed to update the initial state $\spstate^0$.

\vspace{3pt}
\textit{Enclave State Attestation:} 
Step 2 is enclave attestation that needs to be carried out to authenticate the enclave environment including the $\spstate^0$. Note that we add the initial state $\spstate^{0}$ and the public key \pk{} into the enclave measurement $\sigma_{att} = \Sigma$.\sig(\msk{}, (\prog{}, \pk{}, \spstate$^0$)) 
\footnote{Specific implementation may vary depending on the underlying platform.} where $\Sigma$ is a digital signature scheme and \msk{} is the manufacturer-generated secret key for the processor~\cite{pass2017formal}.

The initial state reflects the starting point of \acc{}, which should match the recorded state on the Blockchain. If a node passes the attestation, it means that the \acc{} is set up with the correct on-chain deposit and should be trusted for the subsequent off-chain transactions.

\vspace{3pt}
\textit{Channel Key Establishment:} Once the enclave account is verified, the channel participants start to generate the shared channel key by leveraging any secure two-party key agreement protocols~\cite{bernstein2006curve25519, bresson2007provably}. 

\vspace{3pt}
\textit{Channel Certification:} In this step, an identifier denoted as $\ccid{} := \hash(SORT(\{\pk{}_0, \pk{}_1\}))$ is assigned for the channel, where $\hash$ is a hash function and $SORT$ can be any function used to make sure both parties agree on the same order of \pk{}'s, thus leading to the identical \ccid{}. Next, both ends create a certificate $\cert{}_i:= \Sigma.\sig(\sk{}_i, \pk{}_{1-i})_{i \in \{0, 1\}}$ for the other party by including the target public key \pk{} as the identifier. With the \cert{}, a channel user can claim the \fund{} received from counterpart on the Blockchain when channel is closed.


\vspace{3pt}
\textit{Multi-party State Channel Establishment:} This step is optional for establishing the multi-party state channel. To this end, a group channel-key is generated for securely sharing the channel states among participants. This step cannot complete until after all the necessary two-party channels have been established. Note that the group key only works for the multi-party state channel function and coexists with the keys for direct channels (see Section~\ref{subsub:mpsc}).

\subsection{Key Functions}
\label{subsec:keyfunc}

\vspace{5pt}
\noindent\textbf{Certified Channel:} One main challenge by incorporating TEE into the Blockchain is that current Blockchain implementation does not support remote attestation for TEE platforms. As a result, Blockchain cannot verify the authenticity of the transactional activities from layer-2. To address the problem, we propose \certChannel{} defined below. 

\begin{definition}[\certChannel{}]
A channel in \TheName{} is called \certChannel{} if it is established between two attested enclave accounts and both participants  have the certificate \cert{} issued by the other party.
\label{def:def_auth}
\end{definition}

With the \certChannel{}, Blockchain is agnostic to the enclave attestation and offloads this task to the layer-2 nodes. As long as a node can present a valid certificate issued by the other channel party, Blockchain will trust this enclave node and associated transactions. In this way, the balance security is guaranteed by the attested enclaves. 

\textit{Dispute-free Channels.} The main reason for the disputes existing in prior state channel networks is that it is challenging for Blockchain to discern the old states in an asynchronous network. The victim node may be intentionally blocked to favor the attacker's claim when closing the channel~\cite{lightNn, Raiden, lind2019teechain}. With \certChannel{}, \TheName{} relies on enclaves to correctly update its state before sending out and after receiving transactions. The node locks the channel states if it intends to send a ``claim" transaction to the Blockchain. As a result, channel states are always up to date and the channel can be unilaterally and safely closed without worrying about unstable network connections. In this regard, \TheName{} is free from expensive on-chain dispute resolution operations. 

\vspace{5pt}
\noindent\textbf{Fully Decentralized Channel Network:}
\label{subsub:fdcn}
We envision a fully decentralized channel network (FDCN) will significantly improve the layer-2 network stability, which also aligns with the decentralized design principle of Blockchain. We define a fully decentralized channel network as follows. 

\begin{definition}[Fully Decentralized Channel Network]
A payment/state channel network where a node can establish direct channel connections with other nodes efficiently off the chain and process transactions without relying on intermediaries is a Fully Decentralized Channel Network. 
\label{def:def_fdcn}
\end{definition}

It is economically impossible to turn the current state channel network into an FDCN because it will lock in a significant amount of collaterals in the main chain. \TheName{} addresses this issue by adopting an account-based channel creation to use the on-chain deposit for opening multiple channels off-chain. Another immediate benefit of FDCN is the elimination of intermediaries for transaction routing, thus relieving users from additional fees, operational costs, and security and privacy concerns.

\vspace{5pt}
\noindent\textbf{Multi-Party State Channel:} 
\label{subsub:mpsc}
As discussed in Section \ref{subsec:multiparty}, designing a multi-party state channel is a fundamental challenge but necessary for many off-chain smart contract use cases, such as the multi-party games. Next, we detail our design. 

\vspace{2pt}
\textit{Multi-party channel establishment.}
Before establishing the group channel, we assume that a peer-to-peer channel has already been set up between each pair of members beforehand. With $n$ known participants of a multi-party channel to be created, we first generate the channel id \ccid{} by hashing the sorted public keys of all participants as follows $\ccid{} := \hash(SORT(\{ \pk{}_i \}^{i \in [N]}))$. Then, a group key \gk{} can be generated with any secure multi-party key exchange algorithm~\cite{biswas2008diffie, barua2003extending, boneh2017multiparty}. 


The group key \gk{} is bind with \ccid{}, and only transactions with a tag of the matched \ccid{} can use the key for encryption and decryption. As a result, a transaction in a multi-party channel only needs to be encrypted once and then broadcast to other members.

\begin{figure}[htb]
    \centering
    \includegraphics[width=8cm]{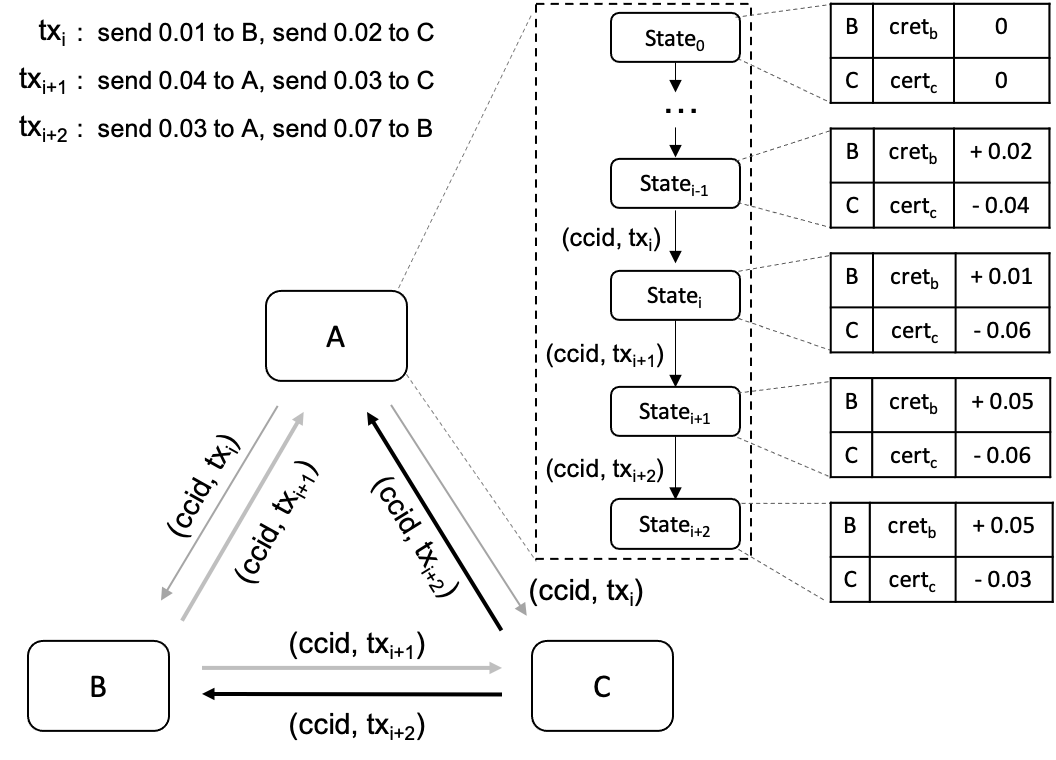}
    \caption{An example of executing a multi-party transfer contract among A, B and C, assuming SORT($\pk{}_A$)$>$SORT($\pk{}_B$)$>$SORT($\pk{}_C$). (+) and (-) in the tables represent the current balance after transactions in each \certChannel{}, respectively.}
    \label{fig:multi}
\end{figure}

\vspace{2pt} 
\textit{Coordinated transaction execution.}
To avoid ambiguity of the transaction execution in a multi-party smart contract scenario, transactions from different parties need to be ordered before processed. In a distributed network, a trusted time source is hard to get for coordination. To address this issue, we let each party $i$ takes turn to send its transactions by the order derived from SORT($\{ \pk_i \}^{i \in [N]}$). Specifically, all the nodes are mute after the channel key is created except for the one with the highest order by the SORT function. Moving forward, all other nodes need to wait for their turn to issue transactions. Figure~\ref{fig:multi} shows an example of the execution of a value-transfer multi-party contract among three-channel members A, B and C. In the figure, \certChannel{}s are opened between any two member nodes. The nodes send transactions \texttt{tx$_i$}, \texttt{tx$_{i+1}$}, and \texttt{tx$_{i+2}$} in the multi-party state channel identified by \ccid{} successively. The figure also shows the balance change of A with other two channel members after each round of communication. Note that the total balance of underlying \certChannel{}s at any time should not surpass the allocated amount by the node for this multi-party channel. Moreover, channel members are also relieved from the concerns of disputes thanks to the unsynchronized state, which is inherited from the underlying \certChannel{}s.



\section{\protSpeedster{} Protocol}

In this section, we first introduce the abstracted ideal functionalities in \TheName{} and then formally present the concrete \TheName{} protocol \protSpeedster{}.

\subsection{Ideal Functionality}
\label{abstraction}
In \protSpeedster{}, two ideal functionalities are assumed: a Blockchain abstraction \idealBC{}$[Contract]$ and a TEE abstraction \idealGatt{} formally defined in \cite{pass2017formal}. As a result, the design and security of \TheName{} are independent of the specific Blockchain and TEE implementations as long as they can provide the required functions. Specifically, we define \idealBC{}$[Contract]$ as an ideal functionality that models the behavior of Blockchain. \idealBC{} defines a smart-contract enabled append-only ledger. The parameter $Contract$ is the smart contract function of the Blockchain. \idealBC{} has an internal $Storage$ that contains the Blockchain data associated with transaction IDs. To append a transaction to the Blockchain, a user sends a transaction to \idealBC{}, which will subsequently trigger the function \sffont{``append"} to execute the transaction (see Figure~\ref{fig:blockchain} in the Appendix for details).

\idealGatt{}~\cite{pass2017formal} provides an abstraction for the general-purpose TEE-enabled secure processor. During initialization, \idealGatt{} creates a key pair as the manufacture key (\msk, \mpk), while \msk{} is preserved in the processor and the \mpk{} could be accessed through \sffont{``getpk"} command. In such an ideal functionality, user first creates an enclave, and loads \prog{} into enclave by sending an \sffont{``install"} command. To call the functions in \prog{}, user sends \sffont{``resume"} command to \idealGatt{} along with the parameters. All operations through the \sffont{``resume"} command of \idealGatt{} is signed with \msk{} by default to ensure the authenticity, whereas \certChannel{} leverages symmetric-key authenticated encryption instead of digital signatures. Therefore, we add a \sffont{switch} to \sffont{``resume"} command to be able to turn off the signature and only when the \sffont{switch} is set, execution output through ``resume" is signed.
(see Figure~\ref{fig:gatt} in the Appendix for detail).

\begin{figure}[hbtp]
\begin{center}
  \textbf{Protocol} $\Pi_{\TheName{}}(\idealP{}_{\mathcolor{0}}, \idealP{}_{\mathcolor{1}}, \idealP{}_{\mathcolor{2}}, ... ,\idealP{}_{\mathcolor{N}}]$
\end{center}
\begin{framed}

\begin{flushleft}

\begin{center}
Program \prog{}
\end{center}


\noindent \textbf{Initially:}

\sffont{bal} := $\emptyset$,
\sffont{certs} := $\emptyset$,
\sffont{channels} := $\emptyset$,
\sffont{state}$^0$ := $\perp$

\bigskip

\setlength\parindent{15pt}

\noindent \colorgray{(1)} \textbf{On receive}(``init")

$(\pk,\sk) \sample \kgen(\secparam)$ \texttt{\textcolor{gray}{ // generate acc$_{enclave}$}}

$\mpk := \mathcal{G}_{att}$.getpk()

\textbf{return} ($\pk, \mpk$)

\bigskip

\noindent \colorgray{(2)} \textbf{On receive} (``deposit", \tx)

parse \tx{} as (\_, \pk', \$val, $\sigma$)

assert \$val $\geq 0$; assert $\sffont{Verify} (\pk, \tx, \sigma)$

\sffont{bal} += \$val; add \tx{} to \spstate{}$^0$

\bigskip

\noindent \colorgray{(3)} \textbf{On receive} (``open", \cid, $\idealR{}$, \inp)

$\ccid := \hash(SORT\{\pk_{\idealR{}}, \pk\})$

abort if \sffont{channels}[\ccid] $\neq \perp$ 

$\ck \sample \bin^*$ \texttt{\textcolor{gray}{// channel key}}

$\cp{} := \{\pk, \pk_{\idealR{}}\}$

(\sffont{st}, \outp{}) := Contract$_{\cid{}}(\sk{}, \text{\sffont{bal}}, \overrightarrow{0}, \cp{})$

append ($\ccid, (\ck, \cid, \sffont{st}, \cp)$) to \sffont{channels}

$\sigma = \sffont{sign} (\sk, {\pk}_R\concat \inp\concat \spstate^0)$;
\cert{} = $(\pk\concat \pk_R\concat \inp\concat \sigma)$


\textbf{return} (\cert, \spstate{}$^0$, \outp{})

\bigskip

\noindent \colorgray{(4)} \textbf{On receive} (``openMulti", \cid, \{\ccid\}$^*$) 

for each $\ccid' \in \{\ccid\}^*$:
\setlength\parindent{25pt}

assert \sffont{channels}[$\ccid'$] $\neq \perp$

extract $\pk{}'$ from \sffont{channels}[$\ccid'$]

\setlength\parindent{15pt}

$\cp := \{\{ \pk'\}^*\cup\pk\}$; $\ccid{} := \hash(SORT(\cp))$

assert \sffont{channels}[\ccid] = $\perp$

$\sffont{gk} \sample \bin^*$ \texttt{\textcolor{gray}{//Group key}}

(\sffont{st}, \outp) := Contract$_{\cid}(\sk_i, \spstate, \overrightarrow{0}, \cp)$

append ($\ccid, (\sffont{gk}, \cid, \sffont{st}, \cp)$) to \sffont{channels}

\ct{} := \sffont{Enc} (\sffont{gk}, \outp{})

\textbf{return} (\ct{})

\bigskip

\noindent \colorgray{(5)} \textbf{On receive} (``authenticate", \ccid, \idealR{}, \cert)

abort if \sffont{certs}[\ccid][\pk$_{\idealR{}}$] $\neq \perp$

parse \cert{} as ($\msg,\sigma$), \sffont{Verify} $(\pk_{\idealR{}}, \msg, \sigma)$

extract $\spstate^0$' from \msg{}, check $\spstate^0$' on Blockchain

 \sffont{certs}[\ccid][\pk$_{\idealR{}}$] := \cert


\bigskip

\noindent \colorgray{(6)} \textbf{On receive} (``send", \ccid, \inp):

assert \sffont{certs}[b] $\neq \perp$

(\ck, \cid, \sffont{st}, \cp) := channles[\ccid]

(\sffont{st\'}, \outp{}) := Contract$_{\cid}(\sk, \spstate, \sffont{st}, \inp)$

update \sffont{channels}[\ccid] to (\ck, \cid, \sffont{st\'}, \cp)

\msg{} := (\pk\concat \inp\concat \sffont{st}'\concat \outp);
\ct{} := \sffont{Enc} (\ck, \msg) 

\textbf{return} (\ct)

\bigskip

\noindent \colorgray{(7)} \textbf{On receive} (``claim")

freeze \textbf{send} function

\tx{} := \{\cert\}$^*$\concat\spstate; $\sigma$ := \sffont{Sign} (\sk,\tx)

\textbf{return} (\tx\concat$\sigma$)



\end{flushleft}

\end{framed}

\caption{\prog{} program of $\Pi_{\TheName{}}$} 
\label{fig:prog_TEE_short}
\end{figure}
\begin{figure}[htbp]

\begin{center}
  \textbf{Protocol} $\Pi_{\TheName{}}(\idealP{}_{\mathcolor{0}}, \idealP{}_{\mathcolor{1}}, \idealP{}_{\mathcolor{2}}, ... ,\idealP{}_{\mathcolor{N}}]$
\end{center}
  
\begin{framed}
\begin{flushleft}

\begin{center}
\contract{}
        
\end{center}
\noindent \textbf{Parameters:}

$Ledger:$ Append only public ledger of \idealBC{}

$Coin:$ Blockchain function that convert value into coins.

\vspace{5pt}
\setlength\parindent{5pt}
\noindent \textbf{On receive} (``deposit", \tx) from \idealP{}:

assert $\tx{} \notin Ledger$

execute \tx{} on the $Blockchain$

append \tx{} to $Ledger$

\bigskip

\noindent \textbf{On receive} (``claim", \tx) from \idealP{}:

parse \tx{} as (\{\cert\}$^*$, \spstate) 
\texttt{\textcolor{gray}{// \spstate{} contains channel data}}

For each \sffont{cert} in \{\cert\}$^*$:

\setlength\parindent{15pt}
    parse \cert{} to (\sffont{to\'}, \sffont{from\'}, $\sigma$ )
    
    abort if $\sffont{Verify} (\sigma, \cert)$ fails \texttt{\textcolor{gray}{// verify the cert}}
    
    extract $\$val$ from $\spstate{}[\sffont{from}]$
    
    assert $\$val \neq 0$ and \sffont{to\'} $=$ \idealP{}

    send(\sffont{from}, \idealP{}, Coin($\$val$)) if $\$val > 0$
    
    send(\idealP{}, \sffont{from}, Coin($-\$val$)) otherwise
    
\setlength\parindent{5pt} 
   
append(\tx) to $Ledger$

\bigskip

\noindent \textbf{On receive} (``read", \tx) from \idealP{}:

output $Ledger[\tx]$

\end{flushleft}
\end{framed}
\caption{On-chain smart contract \contract{} of \protSpeedster{}} 
\label{fig:prog_contract}

\end{figure}
\subsection{\TheName{} Protocol \protSpeedster{}}
\label{subsec:protocol}

We formally present the protocol \protSpeedster{} into two parts: the program \prog{}, in Figure~\ref{fig:prog_TEE_short}, that runs the enclave
and the smart contract \contract{} running on the Blockchain, shown in Figure~\ref{fig:prog_contract}.
In the protocol, we let $\mathcal{P}$ denotes a user, $\mathcal{R}$ as the counterpart users in a channel, and \tx{} represents an on-chain transaction. To execute an off-chain smart contract in \prog{}, we define a function Contract$_{\cid}(\cdot)$ which is parameterized with a smart contract id \cid{}. Contract$_{\cid}(\cdot)$ consumes the channel state and node balance to ensure the balance consistent across channels. Contract$_{\cid}(\cdot)$ generates output \outp{} based on the input and updates the channel state.

\vspace{3pt}
\textit{Node Initialization:}
For the first time to boot up a \TheName{} node, a node sends the \sffont{``install"} command to the enclave to load \prog{}. Then, the node calls the function \colorgray{(1)} of \prog{} by sending message (``init") to create an enclave account \acc{} with key pair (\sk, \pk).
For attestation purpose, a measurement $\sigma_{att}$ of the enclave is generated with the \prog{}, the public key \pk{} of \acc{}, and the node initial state $\spstate^0$.

\vspace{3pt}
\textit{Deposit:}
To deposit, a node needs to complete the following steps. First, it sends a \tx{} to \contract{} on the main chain. The transaction includes the \pk{} of the enclave account \acc{} as the account address. Next, the node  calls the the function \colorgray{(2)} of \prog{} by sending message \sffont{``deposit"} and passing \tx{} as a parameter. Finally, \prog{} verifies the signature of \tx{}, and updates the local initial state $\spstate^0$.

\vspace{3pt}
\textit{Certified Channel:} Each certified channel in \protSpeedster{} is identified by a channel ID $\ccid{}$. A shared channel key $\ck$ is produced in this step. The certificate \cert{} of the channel is created using public keys of both parties. To prevent rollback attacks on $\sigma_{att}$, \prog{} generates a signature $\sigma_{att}$ by signing the tuple (\spstate$^0$, $\{\pk{}_i\}^{i \in \{0,1\}}$, \prog{}) for each channel after function \colorgray{(3)} returns. The tuple is signed under the manufacture secret key \msk{} to reflect the root trust of the hardware. The \cert{} is verified in function \colorgray{(5)}. 

\vspace{3pt}
\textit{Multi-Party State Channel:}
A multi-party state channel is built upon the existing certified channels. To create a multi-party state channel, a node calls the function \colorgray{(4)} of \prog{} by sending \sffont{``openMulti"} message and passes a set of \ccid{} to inform the underlying certified channels with other nodes. We abstract out the process of multi-party shared key generation, which could be replaced by any secure multi-party key negotiation protocol \cite{biswas2008diffie, barua2003extending, boneh2017multiparty}.

\vspace{3pt}
\textit{Transaction:}
To send a channel transaction, a node calls the function \colorgray{(6)} of \prog{} by sending \sffont{``send"} command to \prog{} through \idealGatt{}.resume($\cdot$) and passing the target \ccid{} along with other necessary parameters in \inp{}. Then, \prog{} executes \inp{} with the associated contract by calling Contract$_{\cid{}}(\cdot)$ and updates the channel state accordingly. A channel transaction is constructed over the public key of \pk, the new channel state $\spstate{}'$, the input \inp{}, and the output \outp{}. Then, the transaction is encrypted by an authenticated encryption scheme,such as \GCM{}, with the channel key \ck{}.

\vspace{3pt}
\textit{Claim:}
To claim the \fund{} that $\mathcal{P}$ has received from the channel transactions, the node issues a \sffont{``claim"} call to the function \colorgray{(7)} of \prog{}. \prog{} first freezes all two-party channels, and extracts all the \cert{}s from those channels. The \cert{}s and the local node state \spstate{} constitute the claim transaction \tx{}. \prog{} then signs \tx{} with the private key \sk{} of \acc{} and returns the signed transaction that is further forwarded by the node to the \contract{}. In the end, \contract{} verifies and executes the claim transaction on the Blockchain to redeem the fund for the node. 

\section{Security and Privacy Analysis}
\label{subsec:security}

We formalize a Universal Composability (UC)~\cite{canetti2001universally, badertscher2017bitcoin, kosba2016hawk, lind2019teechain} ideal functionality \idealMain{} (shown in Figure~\ref{fig:Ideal_Main} in the Appendix) to realize the security goals of \protSpeedster{}. 

Participants of \idealMain{} are denoted as $\mathcal{P}$. The internal communication among participants is protected through an authenticated encryption scheme. Following~\cite{canetti2001universally}~\cite{cheng2018ekiden}, we parameterize \idealMain{} with a leakage function $\ell(\cdot):\bin^* \xrightarrow{} \bin^*$ to demonstrate the amount of privacy leaked from the message that is encrypted by the authenticated encryption scheme. The security of \protSpeedster{} is given in Theorem~\ref{th:protocol}.

\begin{theorem}[UC-Security of \protSpeedster{}] If the adopted authenticated encryption \ASE{} is \indcca{} secure and digital signature scheme $\Sigma$ is EU-CMA secure, then the protocol \text{\protSpeedster{}} securely UC-realizes the ideal functionality \text{\idealMain} in the (\idealGatt{}, \idealBC{})-hybrid model for static adversaries.
\label{th:protocol}
\end{theorem}

As defined in Theorem~\ref{th:protocol}, we now formally present the proof that the protocol \protSpeedster{} securely UC-realizes ideal functionality \idealMain{} by showing that an ideal world simulator \idealS{} can simulate the behavior of a real-world adversary \idealA{}.
The security of \protSpeedster{} is proved by showing that \idealS{} can indistinguishably simulate the behavior of \idealA{} for all environment \idealE{}~\cite{canetti2001universally}.

\begin{proof} Let \idealE{} be an environment and \idealA{} be a real-world probabilistic polynomial time adversary~\cite{canetti2001universally} who simply relays messages between \idealE{} and dummy parties. To show that \protSpeedster{} UC-realizes \idealMain{}, we specify below a simulator \idealS{} such that no environment can distinguish an interaction between \protSpeedster{} and \idealA{} from an interaction with \idealMain{} and \idealS{}. That is, for any \idealE{}, \idealS{} should satisfy

\begin{center}
$
    \forall\idealE{}. \text{EXEC}^{\idealE{}}_{\text{\protSpeedster{}, \idealA{}}} \approx \text{EXEC}^{\idealE{}}_{\text{\idealMain{}, \idealS{}}}
$
\end{center}

\textbf{Construction of \idealS{}:}  Please refer Appendix~\ref{apendix:consS} for the construction detail.

\textbf{Indistinguishability: } We show that the execution of the real-world and ideal-world is indistinguishable for all \idealE{} from the view of \idealA{} by a series of hybrid steps that reduce the real-world execution to the ideal-world execution. 

\vspace{3pt}
\noindent $\bullet$ \text{Hybrid} $H_0$ is the real-world execution of \TheName{}.

\vspace{3pt}
\noindent $\bullet$ \text{Hybrid} $H_1$ behaves the same as $H_0$ except that \idealS{} generates key pair (\sk{}, \pk{}) for digital signature scheme $\Sigma$ for each dummy party \idealP{} and publishes the public key \pk{}. Whenever \idealA{} wants to call \idealGatt{}, \idealS{} faithfully simulates the behavior of \idealGatt{}, and relay output to \idealPi. Since \idealS{} perfectly simulates the protocol, \idealE{} could not distinguish $H_1$ from $H_0$.

\vspace{3pt}
\noindent $\bullet$ \text{Hybrid} $H_2$ is similar to $H_1$ except that \idealS{} also simulates \idealBC{}. Whenever \idealA{} wants to communicate with \idealBC{}, \idealS{} emulates the behavior of \idealBC{} internally. \idealE{} cannot distinguish between $H_2$ and $H_1$ as \idealS{} perfectly emulates the interaction between \idealA{} and \idealBC{}, 

\vspace{3pt}
\noindent $\bullet$ \text{Hybrid} $H_3$ behaves the same as $H_2$ except that: If \idealA{} invokes \idealGatt{} with a correct install message with program \prog{}, then for every correct ``resume" message, \idealS{} records the tuple (\outp{}, $\sigma$) from \idealGatt{}, where \outp{} is the output of running \prog{} in \idealGatt{}, and $\sigma$ is the signature generated inside the \idealGatt{}, using the \sk{} generated in $H_1$. Let $\Omega$ denote all such possible tuples. If $(\outp{}, \sigma) \notin \Omega$ then \idealS{} aborts, otherwise, \idealS{} delivers the message to counterpart.
 $H_3$ is indistinguishable from $H_2$ by
reducing the problem to the EUF-CMA of the digital signature
scheme. If \idealA{} does not send one of the correct tuples
to the counterpart, it will fail on attestation. Otherwise, \idealE{} and \idealA{} can be leveraged to construct an adversary that succeeds in a signature forgery.

\vspace{3pt}
\noindent $\bullet$ \text{Hybrid} $H_4$ behaves the same as $H_3$ except that \idealS{} generates a channel key $\ck{}$ for each channel. When \idealA{} communicates with \idealGatt{} on sending transaction through channel, \idealS{} records $\ct{}$ from \idealGatt{}, where $\ct{}$ is the ciphertext of encrypted transaction, using the \ck{} of that channel. Let $\Omega$ denote all such possible strings. If $\ct{} \notin \Omega$ then \idealS{} aborts, otherwise, \idealS{} delivers the message to counterpart. $H_4$ is indistinguishable from $H_3$ by
reducing the problem to the IND-CCA of the authenticated encryption scheme. As \idealA{} does not hold control of \ck{}, it can not distinguish the encryption of a random string and $\Omega$.

\vspace{3pt}
\noindent $\bullet$ \text{Hybrid} $H_5$ is the execution in the ideal-world. $H_5$ is similar to $H_4$ except that \idealS{} emulates all real-world operations. As we discussed above, \idealS{} could faithfully map the real-world operations into ideal-world execution from the view of \idealA{}. Therefore, no \idealE{} could distinguish the execution from the real-world protocol \protSpeedster{} and \idealA{} with \idealS{} and \idealMain{}. 

\end{proof}




Theorem~\ref{th:protocol} also implies stronger privacy protection compared to conventional payment/state channel networks in that: (1) All channels in \TheName{} are created directly between participants. No intermediate node is required to relay transactions, thus alleviating the privacy concerns introduced by HTLC~\cite{green2017bolt, malavolta2017concurrency, dziembowski2019perun, herrera2019difficulty}; (2) the transaction in off-chain channels is encrypted by \GCM{}, and only the enclaves of participants can decrypt it. Therefore, transaction confidentiality is ensured by \TheName{}. 

\vspace{3pt}
\textbf{Preventing Double-Spending Attacks.}
\label{subsec:doublespend}
Each processor has a unique built-in key hard-coded in the CPU~\cite{AMDSEV, sgx:white2} to differentiate their identities during attestation. Moreover,  the processor generates and assigns each enclave a unique identifier~\cite{sgx:white3, AMDSEV} ensuring that even enclaves created by the same processor are distinctive. To prevent the double-spending attacks in the channel, \prog{} updates the balance before sending transactions to the peers. Once the state is updated, it can not be rolled back. Therefore, no fund could be spent multiple times in \TheName{}.

\vspace{3pt}
\textbf{Defending Against TEE Attacks.}
\label{subsec:teeattack}
The hardware-assisted TEE serves as a way to replace complex software-based cryptographic operations. Promising as it is, recent research shows that TEE implementations on specific platforms are vulnerable to the side-channel attacks~\cite{van2018foreshadow, schwarz2017malware, gotzfried2017cache, weichbrodt2016asyncshock, murdock2020plundervolt}, rollback attacks~\cite{costan2016intel, matetic2017rote, brandenburger2017rollback} and incorrect implementation and configurations~\cite{ning2019understanding, buhren2019insecure, intel2019Voltage, beniamini2017trust}. In \TheName{}, we use the TEE abstraction without relying on a specific platform. The design is generalized for different environments and the security has been proven in Theorem~\ref{th:protocol}. On the other hand, we offer suggestions and proactively implement mitigations for the above vulnerabilities from both hardware and software levels. For example, we use SEV-SE~\cite{AMDSEV} to protect against specific speculative side-channel attacks and TCB rollback attacks. We also update the microcode of Intel/AMD/ARM TEE to the latest version. Besides, proper implementation of the system can also help mitigate known side-channel vulnerabilities~\cite{sidechannelmitigate}. \TheName{} uses a side-channel-attack resistant cryptographic library~\cite{mbedtls}, and requires that all nodes are running on the latest version of the firmware to defend against known TEE attacks. Further, an adversary may launch a DoS attack against the node by blocking the Internet connection of the victim or abruptly shut down the OS to force stop the enclave functions.
These DoS attacks are not the focus of this work but can be addressed by adopting a committee enforcement design  \cite{cheng2018ekiden, lind2019teechain}. The state of a channel node is jointly managed by a committee of TEE nodes to tolerate Byzantine fault. Despite the inevitable performance loss in light of the complexity of the committee chain, \TheName{} still outperforms the prior work by enabling efficient multi-party state processing and management in a fully decentralized fashion (see Section~\ref{subsec:keyfunc} and Section~\ref{subsec:channelcomparison}).


\section{Implementation and Evaluation}

In this section, we first overview the implementation of \TheName{} on various platforms (i.e., Intel, AMD, and ARM). Then we elaborate on the configurations of the underlying platforms and present the evaluation results of \TheName{}.

\subsection{Implementation of \TheName{}}
\label{sec:impl}

We build a Virtual Machine (VM) on top of the open-source C++ developed Ethereum Virtual Machine eEVM~\cite{microsoft2019eEVM}, which allows \TheName{} to run Ethereum smart contracts off-the-shelf.
The cryptographic library used in \prog{} is \texttt{mbedTLS}~\cite{mbedtls}, an open-source SSL library ported to TEE~\cite{microsoft2019OEnclave,fan2017smbedtls}. For this work, we adopt 1) \texttt{SHA256} to generate secret seeds in the enclave and to get the hash value of a claim transaction, 2) \texttt{AES-GCM}~\cite{mcgrew2004galois} to authentically encrypt transactions in the state channel,  and 3) \texttt{ECDSA}~\cite{johnson2001elliptic} to sign the \cert{} and the claim transaction. We also customize the OpenEnclave~\cite{microsoft2019OEnclave} to compile the prototype for Intel and ARM platforms. For AMD SEV, we use VMs as the enclaves to run \prog{}. The host can communicate with the enclave via the socket.
To highlight the advantages of \TheName{}, the performances of a few functions are tested, as discussed below.

\vspace{3pt}
\textit{Direct Transactions (Trade):} This function is implemented in C++ and allows users to directly transfer \fund{} or share messages through channels without calling an off-chain smart contract. Before sending a transaction out, the sender first updates the local enclave state ( e.g., the account balance), and then marks the transaction as ``sent". The communication between the sender and receiver enclaves is protected by \texttt{AES-GCM}. 

\vspace{3pt}
\textit{Instant State Sharing:} We implement an instant state sharing function in C++ to allow a user to create direct channels with other users off-chain. We also remove costly signature operations for transactions and replace it with \GCM{}, thereby significantly reducing the communication overhead and enabling instant information exchange (like high-quality video/audio sharing) while preserving privacy. This is difficult to realize with prior research efforts using asymmetric cryptographic functions~\cite{lightNn, miller2019sprites, lind2019teechain}.

\vspace{3pt}
\textit{Faster Fund Exchange:} We implement a \texttt{ERC20} contract~\cite{vogelsteller2015erc} with $50$ LOC in \textit{Solidity}~\cite{dannen2017introducing} to demonstrate the improved performance of \TheName{} in executing off-chain smart contracts. This can be attributed to the elimination of asymmetric signature operations for its off-chain transactions. 

\vspace{3pt}
\textit{Sequential Contract Execution:} To highlight the performance of \TheName{} in executing sequential transaction contracts, we implement the popular two-party Gomoku chess smart contract with $132$ LOC in \textit{Solidity}. Furthermore, players cannot reuse locked \fund{} until the game ends, thus nullifying all benefits from cheating the system.

\vspace{3pt}
\textit{Parallel Contract Execution:}
Applications that need users to act simultaneously, such as \textit{Rock-Paper-Scissors} (RPS), are not easy to achieve in conventional sequentially structured state channels. \TheName{} supports applications running in parallel, faithfully manages the states, and only reveals the final results to players. We implement a typical two-party RPS game with $64$ LOC in \textit{Solidity} to demonstrate this.

\vspace{3pt}
\textit{Multi-party Applications:}
To test the ability of multi-party off-chain smart contract executions, a Monopoly game smart contract with $231$ LOC in \textit{Solidity} is implemented. In this game, players take turns to roll two six-sided dice to decide how many steps to move forward in turn and how to interact with other players.

\subsection{Evaluation}

\textit{SGX platform:} We test \TheName{} with a quad-core 3.6 GHz Intel(R) E3-1275 v5 CPU~\cite{IntelCPU} with 32 GB memory. The operating system that we use is Ubuntu 18.04.3 TLS with Linux kernel version 5.0.0-32-generic. We also deploy LN nodes~\cite{lnd} as the baseline for comparison on another physical machine with the same configurations.

\vspace{3pt}
\noindent \textit{SEV platform:} We evaluate \TheName{} on an SEV platform with 64 GB DRAM and an SEV-enabled AMD Epyc 7452 CPU~\cite{AMDCPU}, which has 32 cores and a base frequency of 2.35 GHz. The operating system installed on the AMD machine is Ubuntu 18.04.4 LTS with an AMD patched kernel of version 4.20.0-sev~\cite{AMDSEV}. The version of the QEMU emulator that we use to run the virtual machine is 2.12.0-dirty. The virtual machine runs Ubuntu 18.04 LTS with the kernel version 4.15.0-101-generic and 4 CPU cores.

\vspace{3pt}
\noindent \textit{TrustZone platform:} The evaluation of TrustZone is carried out in the QEMU cortex-a57 virtual machine with 1 GB memory and Linux buildroot 4.14.67-g333dc9e97-dirty as the kernel. 

\begin{table}[H]
\caption{Line of code in Speedster.} 
\label{tab:loc}
\centering
\footnotesize
\begin{tabular}{lllcc}
\toprule
 & \multicolumn{1}{c}{\textbf{Component}} & \multicolumn{1}{c}{\textbf{Code}} & \textbf{LOC} & \textbf{Total(\#)} \\ \midrule
\textbf{Shared} & \sffont{eEVM}~\cite{microsoft2019eEVM} & C++ & 25.3k & 25.3k \\ \midrule
\multirow{2}{*}{\textbf{SGX/TrustZone}} & \prog{} & C++ & 3.1k & \multirow{2}{*}{5.4k} \\ \cmidrule(lr){2-4}
 & other & C++ & 2.3k &  \\ \midrule
\multirow{2}{*}{\textbf{AMD SEV}} & \prog{} & C++ & 3.7k & \multirow{2}{*}{7.8k} \\ \cmidrule(lr){2-4}
 & other & C++ & 4.1k &  \\ \bottomrule
\end{tabular}
\end{table}

\subsubsection{Code Size}
To port eEVM into \TheName{}, we added extra $650$ LOC to eEVM. In general, the eEVM contains $3.2k$ LOC in C++ and another $22.1k$ LOC coming from its reliance.

\TheName{} is evaluated on Intel, AMD, and ARM platforms with around $38.5k$ LOC in total, as shown in Table~\ref{tab:loc}. Specifically,  $25.3k$ LOC comes from the contract virtual machine eEVM~\cite{microsoft2019eEVM} which is shared with  all cases. \prog{} has $3.1k$ LOC in C++ for SGX/TrustZone and $3.7k$ LOC for AMD SEV.
The \contract{} deployed on the Blockchain is implemented with $109$ LOC in \textit{Solidity}.

\subsubsection{Time Cost for Transaction Authentication}
\label{subsec:algocomparison}

\begin{figure}[htp]
    \centering
    \includegraphics[width=8.5cm]{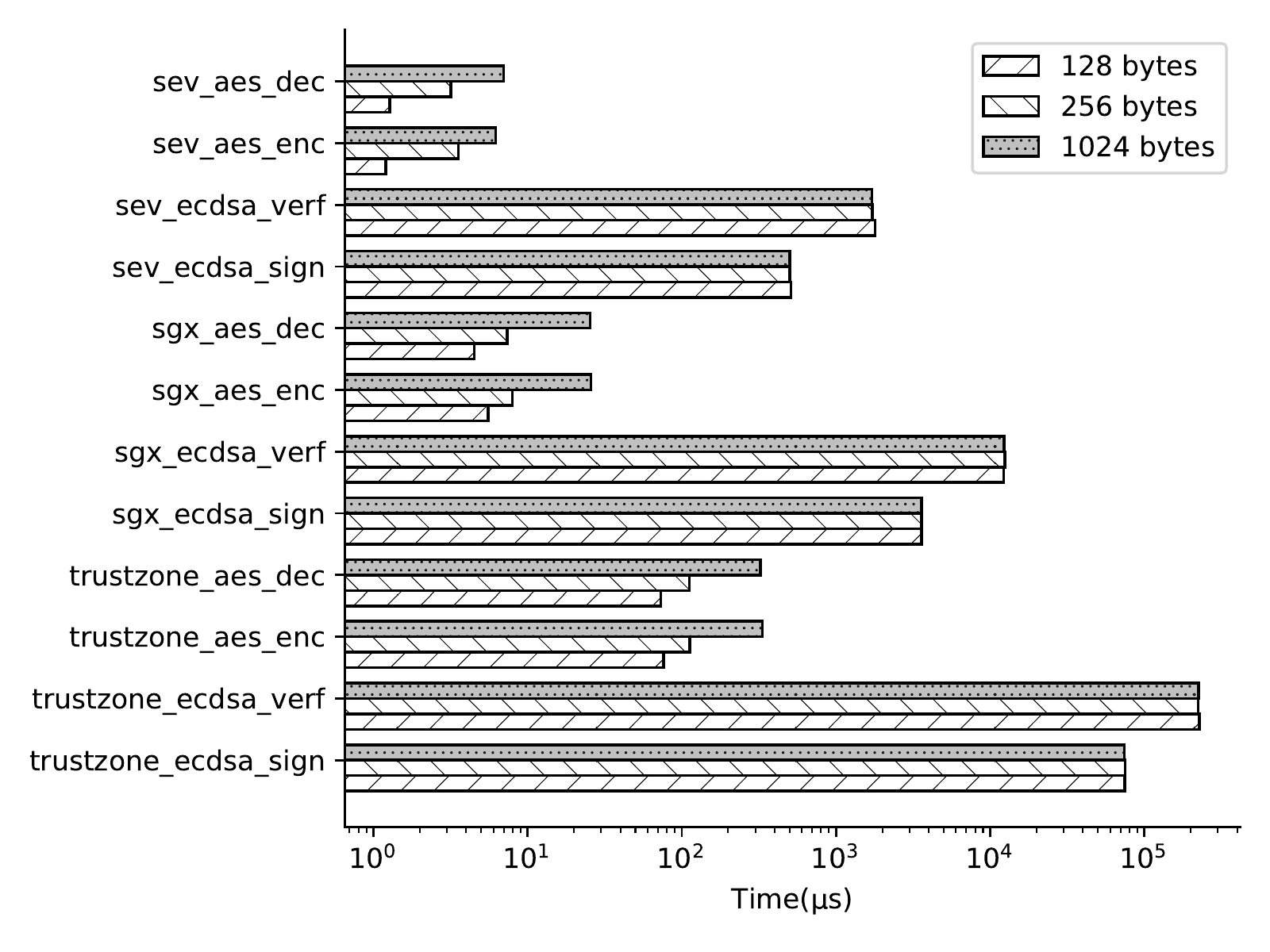}
    \caption{Performance comparison between \ECDSA{} and \GCM{} enabled transaction security on SGX, SEV, and TrustZone platforms.}
    \label{fig:sev}
\end{figure}

In the \TheName{} prototype, we use \GCM{} to replace \ECDSA{} that is adopted in previous channel projects for transaction authentication. By trusting the secure enclave,  \TheName{} uses efficient symmetric operations to realize both confidentiality and authenticity of transactions at the same time. Figure~\ref{fig:sev} shows the comparison of the performance between \ECDSA{} and \GCM{} when processing data of the size $128$ bytes, $256$ bytes, and $1024$ bytes, respectively. This experiment is carried out on Intel, AMD, and ARM platforms with four operations: \ECDSA{} sign, \ECDSA{} verify, \GCM{} encrypt, and \GCM{} decrypt. \ECDSA{} is evaluated under \texttt{secp256k1} curve. The key size of \ECDSA{} is $256$ bits while that of \GCM{} is $128$ bits. 

Figure~\ref{fig:sev} is plotted on a log scale. We can see that regardless of the tested platform, \GCM{} is $3-4$ orders of magnitude faster. Besides, \GCM{} performs better with small-sized messages. With increased data size, the time cost of \ECDSA{} remains constant while that of \GCM{} grows. This is because \ECDSA{} always signs a constant hash digest rather than the actual data. In practice, the average transaction size on the Ethereum is $405$ bytes~\cite{ave2020txsize}. Therefore, using symmetric-key operations will significantly boost the transaction-related performance.

\subsubsection{Transaction Performance}

We evaluate \TheName{} on the time cost for transactions in a direct channel under the test cases in Section~\ref{sec:impl} on Intel, AMD, and ARM platforms. In this experiment, we use the popular layer-2 network, the LN, as the baseline. We measure the time cost for transactions over a direct channel, which may include the time cost for transaction generation and confirmation, corresponding contract execution, transmission in the local network, and other related activities in a life cycle of an off-chain transaction. We test \TheName{} in \GCM{} mode to reflect our intended symmetric-key design. Additionally, we also test the batching transaction performance as a comparison with TeeChain~\cite{lind2019teechain}.

\begin{table}[htp]
  \centering
\caption{Local time cost for end-to-end transaction ($ms$).}
\label{tab:timecost}
\footnotesize
\begin{tabular}{lllll} 
\toprule
 & \multicolumn{1}{l}{\textbf{Payment}} & \multicolumn{1}{l}{\textbf{ERC20}} & \multicolumn{1}{l}{\textbf{Gomuku}} & \multicolumn{1}{l}{\textbf{RPC}} \\ \midrule
 {LN} & 192.630 & N/A & N/A & N/A \\ 
{SEV:\GCM{}} & 0.1372 & 0.1382 & 0.6667 & 0.1365 \\
{SGX:\GCM{}} & 0.0205 & 0.3500 & 0.4500 & 0.1930 \\ 
{TZ:\GCM{}} & 20.496 & 40.148 & 95.092 & 37.215 \\ \bottomrule
\end{tabular}
\end{table}

The experiment result is an average of 10,000 trials and is shown in Table~\ref{tab:timecost} with the implemented smart contracts ERC20, Gomoku, and rock-paper-scissor (RPC).

\vspace{2pt}
\textit{Evaluation on SGX:}
The evaluation of \TheName{} on the SGX platform is carried out by running two \TheName{} instances on the same SGX machine. Direct transaction without contract execution takes $0.0205 ms$ with \GCM{}, which is four orders of magnitude faster compared to LN. When calling smart contracts, it takes $0.1930 ms - 0.4500 ms$ to process a contract-calling transaction.

\vspace{2pt}
\textit{Evaluation on AMD:}
As there is no available AMD cloud virtual machine that supports SEV, we only evaluate \TheName{} on the AMD platform by running the \prog{} in two Ubuntu guest virtual machines as the enclaves. To protect the code and data of \prog{} that runs in the enclave, we only allow users to access \prog{} by calling the related interface through the socket.

For the direct transaction, SEV:\GCM{} takes an average of $0.1372 ms$. When invoking smart contracts, the time cost varies for different applications. As shown in Table~\ref{tab:timecost}, RPC ($0.1365 ms$) and ERC20 ($0.1382 ms$) are faster than Gomoku ($0.6667 ms$) due to their simple logic and fewer steps to take.

\vspace{2pt}
\textit{Evaluation on ARM:}
As the evaluation of ARM TrustZone runs upon the QEMU emulator, the performance of ARM is the worst. Nevertheless,  the evaluation result in Table~\ref{tab:timecost} shows that \prog{} takes $20.496 ms$ to run direct transactions, which is around $9$ $\times$ better than LN. For the smart contract execution, it takes $30-90 ms$ to process contract transactions.

\begin{table}[ht]
\centering
\footnotesize
\caption{Channel performance.}
\label{tab:performance}
\begin{tabular}{@{}llllll@{}}
\toprule
\multirow{2}{*}{} & \multicolumn{3}{c}{\textbf{Throughput} $(tps)$} & \multicolumn{2}{c}{\textbf{Latency} ($ms$)} \\ \cmidrule(l){2-6} 
 & \multicolumn{1}{c}{LN(lnd)} & \multicolumn{1}{c}{Speedster} & \multicolumn{1}{c}{\begin{tabular}[c]{@{}c@{}}Change\\ ($\times$)\end{tabular}} & \multicolumn{1}{c}{LN(lnd)} & \multicolumn{1}{c}{Speedster} \\ 
Payment & 14 & 72,143 & 5153$\times$ & 548.183 & 80.483 \\ 
ERC20 & N/A & 30,920 & N/A & N/A & 82.490 \\
RPC & N/A & 53,355 & N/A & N/A & 80.743 \\ 
Gomoku & N/A & 2,549 & N/A & N/A & 82.866 \\ \bottomrule
\end{tabular}
\end{table}

\begin{figure}[htp]
    \centering
    \includegraphics[width=8.5cm]{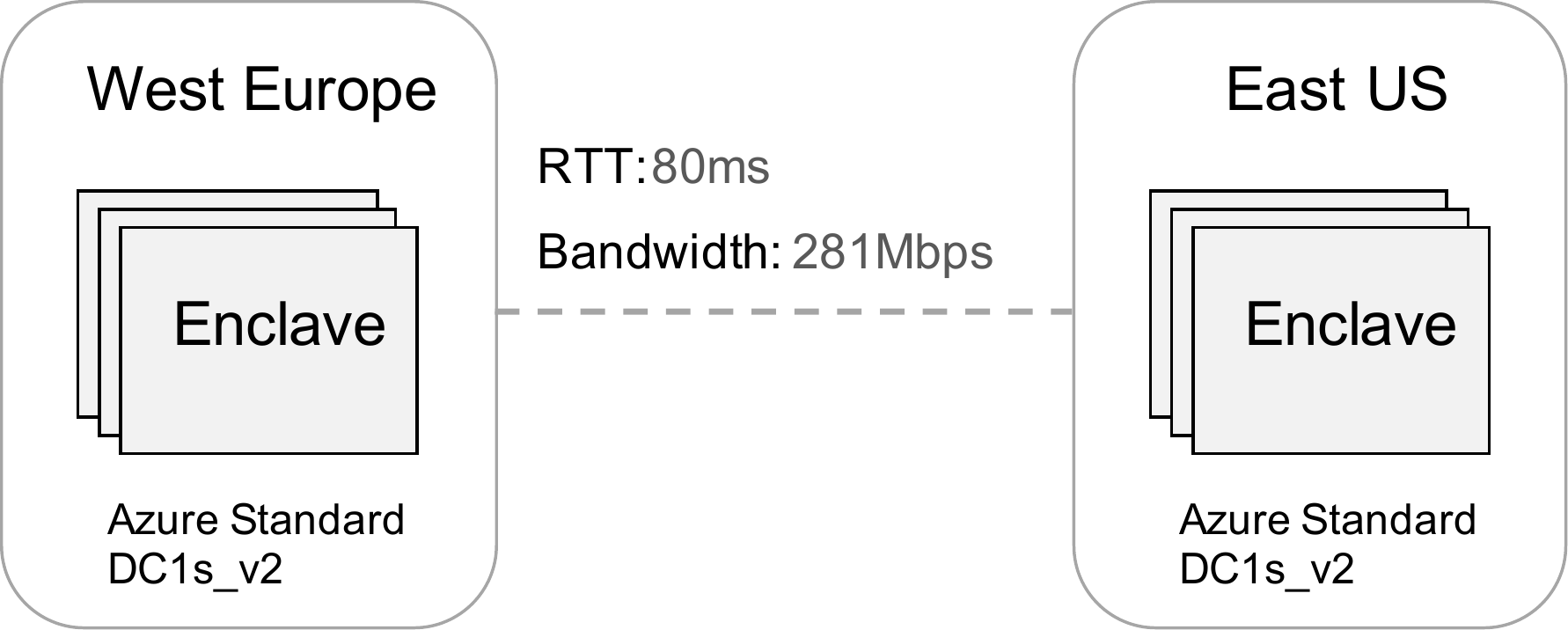}
    \caption{Network setup for the evaluation.}
    \label{fig:network_sgx}
\end{figure}

\vspace{2pt}
\textit{Real-world Evaluation:}
\label{subsec:realeval}
To evaluate the performance of \TheName{} in the real world, we deploy \TheName{} on two Azure Standard DC1s\_v2 (1 vCPUs, 4 GB memory) virtual machines, which are backed by the 3.7GHz Intel XEON E-2288G processor, one in East US, and the other in West Europe, as shown in Figure~\ref{fig:network_sgx}. The kernel of the virtual machine is 5.3.0-1034-azure, and the operating system is version 18.04.5 LTS. LN node is deployed and evaluated on the machine as a baseline to highlight the significant performance improvement of \TheName{}. Table~\ref{tab:performance} shows the evaluation result. The throughput of LN is $14 tps$ while \TheName{} achieves $72,143 tps$ on payment operation, $5,000\times$ more efficient than LN. Specifically, the latency to execute a \TheName{} transaction is around $80 ms$, close to the RTT between testing hosts, while the latency to run an LN payment transaction is around $500 ms$. 

TeeChain is a TEE-supported payment channel network \cite{lind2019teechain}. We tried hard to run a head-to-head comparison with it but failed to do so
\footnote{Though TeeChain is open source, we were not able to successfully run the project even after we contacted the author of TeeChain.}. Instead, we provide insights for a theoretical comparison. TeeChain nodes coupled with committee chains to defend against node failure. \TheName{} can be adapted to a similar design but inevitably sacrifices the performance
\footnote{Each fund spending needs to be approved by the committee using a multi-signature.}. In this regard, the throughput of the committee-based \TheName{} will be comparable with that of TeeChain. However, \TheName{} is much more efficient in off-chain channel creation/closure (see Section~\ref{subsec:cost}
) and supports multi-party state processing.

\subsubsection{Channel System Comparison}
\label{subsec:channelcomparison}

\begin{table*}[t]
\centering
\caption{Feature comparison with other channel projects}
\label{tab:func}
\footnotesize
\begin{tabular}{@{}lcccccccc@{}}
\toprule
\multirow{2}{*}{\textbf{Features}} & \multicolumn{8}{c}{\textbf{Channel Projects}} \\ \cmidrule(l){2-9} 
 &  LN~\cite{lightNn} & TeeChan~\cite{lind2016teechan} & TeeChain~\cite{lind2019teechain} & DMC~\cite{decker2015fast} & SFMC~\cite{burchert2018scalable} & Perun~\cite{dziembowski2019perun} & Celer~\cite{dong2018celer} & Speedster  \\ \midrule
Direct Off-chain Channel Open & \xmark{} & \cmark{} & \cmark{} & \xmark{} & \cmark{} & \cmark{} & \xmark{} & \cmark{} \\ \midrule
Direct Off-chain Channel Close & \xmark{} & \xmark{} & \cmark{} & \xmark{} & \cmark{} & \cmark{} & \xmark{} & \cmark{} \\ \midrule
Dynamic Deposit & \xmark{} & \xmark{} & \cmark{} & \xmark{} & \cmark{} & \xmark{} & \xmark{} & \cmark{} \\ \midrule
Off-Chain Contract Execution & \xmark{} & \xmark{} & \xmark{} & \xmark{} & \xmark{} & \xmark{} & \cmark{} & \cmark{} \\ \midrule
Fully Distributed & \xmark{} & \xmark{} & \xmark{} & \xmark{} & \xmark{} & \xmark{} & \xmark{} & \cmark{} \\ \midrule
Multi-Party State Channel & \xmark{} & \xmark{} & \xmark{} & \xmark{} & \cmark{} & \xmark{} & \xmark{} & \cmark{} \\ \midrule
Dispute-Free & \xmark{} & \xmark{} & \cmark{} & \xmark{} & \xmark{} & \xmark{} & \xmark{} & \cmark{} \\ \midrule
Duplex Channel & \xmark{} & \xmark{} & \cmark{} & \cmark{} & \cmark{} & \xmark{} & \xmark{} & \cmark{}  \\ \bottomrule
\end{tabular}
\end{table*}

\TheName{} supports efficient multi-party state channel creation and closure. To highlight the advantages of \TheName{}, we compare \TheName{} with other major existing channel projects. Table~\ref{tab:func} shows the difference in terms of the following features: Direct off-chain channel open/closure, dynamic deposit (dynamically adjusting \fund{} in an existing channel on demand~\cite{lind2019teechain}), 
symmetric-key operations for transactions (using symmetric encryption algorithms to ensure the authenticity and privacy of the off-chain transaction), off-chain smart contract execution, 
full decentralization (see Definition~\ref{def:def_fdcn}), 
multi-party state channel, dispute-free, and duplex channel (where both participants of a channel can send \fund{} back and forth).

We compare the functions provided by \TheName{} and TeeChain. TeeChain is not a fully decentralized channel network. Despite the dynamic deposit and bilateral  termination~\cite{lind2019teechain}, every channel opened in TeeChain has to be associated with a deposit locked on the main chain. As a result, similar to the Lightning network, creating many channels requires freezing a significant amount of collateral on the Blockchain and incurring expensive on-chain operations. Therefore, it is not realistic to build direct channels for any pair of nodes in the network. Alternatively, TeeChain still largely depends on HTLC for transaction routing in practice, which leads to privacy concerns. On the contrary, a deposit in \TheName{} can be shared by multiple off-chain channels. Direct channels can be efficiently established. Further, TeeChain does not support the off-chain smart contract execution and multi-party state channels. The pairwise channel structure of TeeChain confines the state within the channel. In contrast, due to balance sharing and \certChannel{}, states across multiple channels can be managed and exchanged authentically in the same \TheName{} account.

In Perun~\cite{dziembowski2019perun}, virtual channels can also be opened and closed off the Blockchain, but once the channel is created, the underlying ledger channels have to be locked. The minimum funds across the ledger channels determine the available capacity. As shown in Table~\ref{tab:func}, \textbf{\TheName{} is the only off-chain state channel project that accomplishes all the listed functions}.

\subsubsection{Main Chain Cost}
\label{subsec:cost}

\begin{table*}[t]
 \centering
\caption{Number of on-chain transactions and Blockchain Costs (BC) per channel.}
\footnotesize
\label{tab:blockchaincost}
\begin{tabular}{lcccccccccc}
\toprule
\multirow{2}{*}{\textbf{\begin{tabular}[c]{@{}l@{}}Payment Channel\end{tabular}}} & \multicolumn{2}{c}{\textbf{Setup}} & \multicolumn{2}{c}{\textbf{Open Channel}} & \multicolumn{2}{c}{\textbf{Close Channel}} & \multicolumn{2}{c}{\textbf{Claim}} & \multicolumn{2}{c}{\textbf{Total}} \\ \cmidrule(l){2-11}
 & No.tx & BC & No.tx & BC & No.tx & BC & No.tx & BC & No.tx & BC \\ \midrule

LN~\cite{lightNn} & 2 & 2 & 1 & 2 & 1 & 2 & N/A & N/A & 4 & 6 \\ \midrule

TeeChain~\cite{lind2019teechain} & 1 & 1+$p$/2 & off chain & off chain & off chain & off chain & 1 & 1+$p$/2+$m$ & 2 & 2+$p$+$m$ \\ \midrule

DMC~\cite{decker2015fast} & N/A & N/A & 1 & 2 & 1 & 2 & N/A & N/A & 2 & 4 \\ \midrule

SFMC~\cite{burchert2018scalable} & 1/$\aveChannel{}$ & p/$\aveChannel{}$ & off chain & off chain & off chain & off chain & 1/$\aveChannel{}$ & $p$/$\aveChannel{}$ & 2/$\aveChannel{}$ & 2$p$/$\aveChannel{}$ \\ \midrule

Speedster & 1/$\aveChannel{}$ & 1/$\aveChannel{}$ & off chain & off chain & off chain & off chain & 1/$\aveChannel{}$ & 1/$\aveChannel{}$ & 2/$\aveChannel{}$ & 2/$\aveChannel{}$ \\ \bottomrule

\end{tabular}
\end{table*}

Similar to the previous work~\cite{burchert2018scalable, lind2019teechain}, we evaluate the main chain cost: (1) the number of required on-chain transactions and (2) the number of pairs of public keys and signatures that are written to the Blockchain (defined as Blockchain cost in~\cite{burchert2018scalable}). 

We select a set of representative channel projects to evaluate and be compared with \TheName{}. In particular, we choose LN~\cite{lightNn} (the most popular payment channel system in reality), DMC~\cite{decker2015fast} (a duplex payment channel), TeeChain~\cite{lind2019teechain} (a TEE-based channel project), and SFMC~\cite{burchert2018scalable} (it also supports off-chain channel open/closure). The comparison is carried out by analyzing each project under bilateral termination~\cite{lind2019teechain}, i.e., a channel is closed without disputes. The result is shown in Table~\ref{tab:blockchaincost}. We take LN and TeeChain, for example, to demonstrate the cost efficiency of \TheName{}. 

Before opening an LN channel, each node has to send one on-chain transaction with a Blockchain Cost (BC) of $1$ to commit a deposit in the channel. Then, each LN channel has to send one on-chain transaction with a BC of $2$. To close this channel, one of the channel's participants needs to send a transaction with the latest channel state and signatures from both sides to the Blockchain. 

In TeeChain~\cite{lind2019teechain}, a group of committee nodes handles and dynamically associates deposits with channels. Thus, at least one ``deposit" transaction is needed to set up the system with a BC of $1+p/2$, where $p$ is the size of the committee. Since TeeChain can also close the channel off the chain, there is no associated cost for that. All committee members in TeeChain use the same $m$-out-of-$p$ multi-signature for each ``deposit" transaction, so the BC is  $1+p/2+m$.

In contrast, a deposit to a \TheName{} can be freely allocated to different channels. Therefore, we only need $1$ 
``deposit" transaction to initialize the account and create \aveChannel{} channels. 
There is no cost for channel opening and closing as \TheName{} can do it completely offline. To claim the remaining \fund{} from active channels, one on-chain transaction needs to be sent. Assuming that one deposit and one claim transactions are shared by $\aveChannel{}$ channels on average, \TheName{} requires $2/\aveChannel{}$ on-chain transactions with a BC of $2/\aveChannel{}$ for each channel on average. 

In summary, we observe that \TheName{} needs $80\%$ less on-chain transactions than LN and the same number of transactions as TeeChain when $\aveChannel{} \geq 2$ and one deposit when a $2$-out-of-$3$ multi-signature is used for each TeeChain channel. For the BC of each channel, \TheName{} outperforms LN by at least $66\%$ when $\aveChannel{} \geq 2$, and $97\%$ if $\aveChannel{} \geq 11$~\cite{ave2019channel}. Compared to TeeChain, \TheName{} reduces over $84\%$ BC when $\aveChannel{} \geq 2$.

\section{Discussion}
In this section, we discuss other aspects of \TheName{}, i.e., partial claim from \acc{}, unavailability of TEE, and cross-chain state channel.

\vspace{3pt}
\textit{Partial Claim}. In \TheName{}, one ``claim" transaction can be used to freeze all the channels that belong to the same user and to withdraw all the \fund{} from these channels. We can also extend the current function to support the partial claim. In other words, the user can decide to claim a portion of the total funds in a channel. As a result, the partially-claimed channel can still operate as normal.

\vspace{3pt}
\textit{Single Node Failure}. Our work is consistent with prior TEE-based research that the availability of TEE is out of the investigation. Nevertheless, we briefly discuss how to ensure the channel balance safety when TEE fails to work correctly, such as power failure and network issues. First, unavailability will not cause a user to lose state. The state of a faulty node can be recovered by authenticated states from other correct nodes. Alternatively, \TheName{} can also adopt committee-based protocols \cite{cheng2018ekiden,lind2019teechain} to manage the account state. Since the committee is only responsible for the channel setup, it does not affect the channel performance, such as throughput and other unique functionalities of \TheName{} compared to prior work, such as TeeChain.


\vspace{3pt}
\textit{Cross-chain State Channel}. Running inside the enclave, \prog{} does not rely on any particular Blockchain platform to operate. Therefore, it is possible to generate multiple enclave accounts and deposit from multiple Blockchain projects to these accounts. Ideally, this would allow \TheName{} to create a cross-chain state channel that could significantly reduce related costs. We will leave this as our future work.


\vspace{3pt}
\textit{Unilateral \certChannel{}}.
The current \certChannel{} is bilateral, which requires both participants to run in TEE-enabled nodes and to issue \cert{}s. This will be a burden to manage \cert{}s for the nodes that intend to receive from other nodes. In the future work, we plan to improve the protocol to enable a user to create unilateral \certChannel{} where only the sender is required to run in the enclave.
\section{Related Work}

\noindent\textbf{HTLC Privacy and Security:} 
HTLC is one of the fundamental building blocks in the current layer-2 channel design to facilitate transactions between parties without direct channel connections \cite{Raiden, lightNn}. But HTLC has privacy issues~\cite{green2017bolt, malavolta2017concurrency, dziembowski2019perun, herrera2019difficulty} and is vulnerable to various attacks~\cite{malavolta2019anonymous, tsabary2020mad, tochner2019hijacking, khalil2017revive}. MAPPCN~\cite{tripathy2020mappcn}, MHTLC~\cite{malavolta2017concurrency}, AMHL~\cite{malavolta2019anonymous}, and CHTLC~\cite{yu2019chameleon} tried to address the privacy issues introduced by HTLC by adding additional countermeasures. MAD-HTLC~\cite{tsabary2020mad} presented the mutual assured destruction HTLC that could mediate the bribery attack. 
Nevertheless, they introduce extra overhead and still require HTLC. Perun~\cite{dziembowski2019perun} enabled a user to create a virtual payment channel to avoid HTLC, but it can only span two ledger channels. In contrast, \TheName{} allows all nodes to connect directly without relying on HTLC and expensive on-chain operations.  

\vspace{2pt}
\noindent\textbf{Efficient Channel Network:} Multi-hop transactions in existing channel networks~\cite{lightNn, Raiden} incurs non-negligible overhead with capacity and scalability issues. Current channel design addresses the problem with distinct focuses. For example, MicroCash~\cite{almashaqbeh2019microcash} introduced the escrow setup that supports concurrent micropayments. 
Sprites~\cite{miller2019sprites} is built on LN and reduced the latency of LN in multi-hop transactions. 
Celer~\cite{dong2018celer} leveraged a provably optimal value transfer routing algorithm to improve the HTLC routing performance. 
Pisa~\cite{mccorry2019pisa} enabled the party to delegate itself to a third party in case it goes off-line. 
REVIVE~\cite{khalil2017revive} rebalanced the fund in channels to increase the scalability of the payment channel network. 
Liquidity Network~\cite{felley2018towards,khalil2018nocust} used hubs to connect users, which raises privacy and centralization concerns. 
\TheName{} is a fully decentralized account-based channel network, which outperforms the existing channel networks. 

\vspace{2pt}
\noindent\textbf{Multi-Party Channel Network:} Several related works offer multi-party payment/state channel solutions. 
Based on Perun, Dziembowski \textit{et al.} proposed the first multi-party state channel~\cite{dziembowski2019multi} that operates recursively among participants. Burchert {et al.}~\cite{burchert2018scalable} presented a multi-party channel with timelocks by adding a new layer between the Blockchain and the payment channel. 
Hydra~\cite{chakravarty2020hydra} introduced an isomorphic multi-party state channel by directly adopting the layer-1 smart contract system.  \TheName{} establishes a multi-party channel directly among participants without intermediaries, thus reducing cost and enhancing security.

\vspace{2pt}
\noindent\textbf{Blockchain projects based on trusted hardware:} Using trusted hardware provides promising solutions to the issues in Blockchain. For instance, Town Crier~\cite{zhang2016town} used SGX to implement authenticated data feed for smart contracts. Ekiden~\cite{cheng2018ekiden} and FastKitten~\cite{das2019fastkitten} proposed Blockchain projects that can elevate the confidentiality of the smart contract. 
In Tesseract~\cite{bentov2019tesseract}, credits could be exchanged across multiple chains.  
Obscuro~\cite{tran2018obscuro} built a privacy-preserving Bitcoin mixer. All these projects are intended for layer-2. 



\vspace{2pt}
For layer 2, TeeChan~\cite{lind2016teechan} was built on top of the Lightning network and created new channels instantly off-chain. However, it still requires synchronization with Blockchain and cannot create multiple channels with a single deposit. Based on TeeChan, TeeChain~\cite{lind2019teechain} was proposed to set up a committee for each node and dynamically allocate the deposit to channels, but it is a payment channel system and still requires HTCL for multi-hop transactions. In contrast, \TheName{} is a fully decentralized multi-party state channel system and provides better privacy protection.  

\section{Conclusion}
\TheName{} is the first account-based state channel system, where off-chain channels can be freely opened/closed using the existing account balance without involving Blockchain. 
\TheName{} introduces \certChannel{} to eliminate the expensive operations for transaction processing and dispute resolution. To the best of our knowledge, \TheName{} is the first channel system that achieves FDCN, thus eliminating the risks and overhead introduced by HTLC once for all. With \certChannel{} and FDCN, \TheName{} is capable of executing multi-party state channel efficiently. The practicality of \TheName{} is validated on different TEE platforms (i.e., Intel SGX, AMD SEV, and ARM TrustZone). The experimental results show much-improved performance compared to LN and other layer-2 channel networks.

\balance

\bibliographystyle{ACM-Reference-Format}

\bibliography{main}

\newpage
\begin{appendix}
\label{sec:appendix}

\section{Construction of $\idealS{}$}
\label{apendix:consS}
 \idealS{} simulates \idealA{}, \idealMain{} internally. \idealS{} forwards any input $e$ from \idealE{} to \idealA{} and records the traffic going to and from \idealA{}.

\vspace{3pt}
\noindent (1) \textit{Deposit:} If \idealPi{} is honest, \idealS{} obtains (``deposit", \tx{}) from \idealMain{}, and emulates a call of ``deposit" to \idealGatt{} through ``resume" interface. Otherwise, \idealS{} reads \tx{} from \idealE{} and emulates message (``deposit", \tx{}) to \idealMain{} with the identity of \idealPi{}, then sends the  ``deposit" call to \idealGatt{}.

\vspace{3pt}
\noindent (2) \textit{Open Channel:} When \idealPi{} is honest, \idealS{} emulates a call of ``open" to \idealGatt{} on receiving (``open",\ccid{}, \cid{}, \idealPi{}, \idealPj{}) from \idealMain{}.

When  \idealPi{} is corrupted:
\begin{itemize}
    \item \idealS{} obtains a public key \pk, and a smart contract id \cid{} from \idealE{}, then generate a random string as \inp{}. \idealS{} sends the message (``open", \cid{}, $\idealR{}, \inp{}$) to \idealMain{} and collect the output with the identity of \idealPi{}. Then \idealS{} emulates a ``resume" call to \idealGatt{} with the same messages (``open" , $\cid{}, \idealR{}, \inp{}$) on behalf of \idealPi{} and collect the output from \idealGatt{}.
    \item Upon receiving (``open", \ccid{}, \cid{}, $\idealPi{}, \idealPj$) from \idealMain{}. \idealS{} obtains \inp{} from \idealE{} and emulates a ``resume" call to \idealGatt{} sending message (``open", \cid{}, \idealPj{}, \inp{}) on behalf of \idealPi{} and record the output from \idealGatt{}
\end{itemize}
\vspace{3pt}
\noindent (3) \textit{Channel Authentication:}  Upon receiving message (\idealPi{}, \idealPj{}, ``authenticate", \cert{}) from \idealMain{} of an honest \idealPi{}, \idealS{} records \cert{}. \idealS{} emulates a ``resume" call to \idealGatt{} sending message (``authenticate", \ccid{}, \idealPj{}, \cert{}). Then, \idealS{} sends an ``OK" command to \idealMain{}.

If \idealPi{} is corrupted, \idealS{} obtains a public key \pk, a channel id\ccid{} from \idealE{}, a \sk{} from a signature challenger \sffont{SCh}, then generates a random string as $m$. \idealS{} computes $\sigma := \Sigma.\sig(\sk{}, \pk{}\concat m)$, then sends the message (``authenticate", \pk{}, \ccid{}, $(\pk_i\concat\pk\concat m\concat\sigma)$) to \idealMain{} and collect the output with the identity of \idealPi{}. Then \idealS{} emulates a ``resume" call to \idealGatt{} with the same messages on behalf of \idealPi{} and collect the output from \idealGatt{}.
\vspace{3pt}
\noindent (4) \textit{Multi-party Channel:} Upon receiving message (``openMulti", \ccid{}, \cid{}, \{\ccid{}\}$^*$) from \idealMain{} of an honest \idealPi{}, \idealS{} emulates a ``resume" call to \idealGatt{} sending message (``openMulti", \cid{}, \{\ccid\}$^*$). Then relay the output to \idealPi{}.

\begin{figure}[htbp]
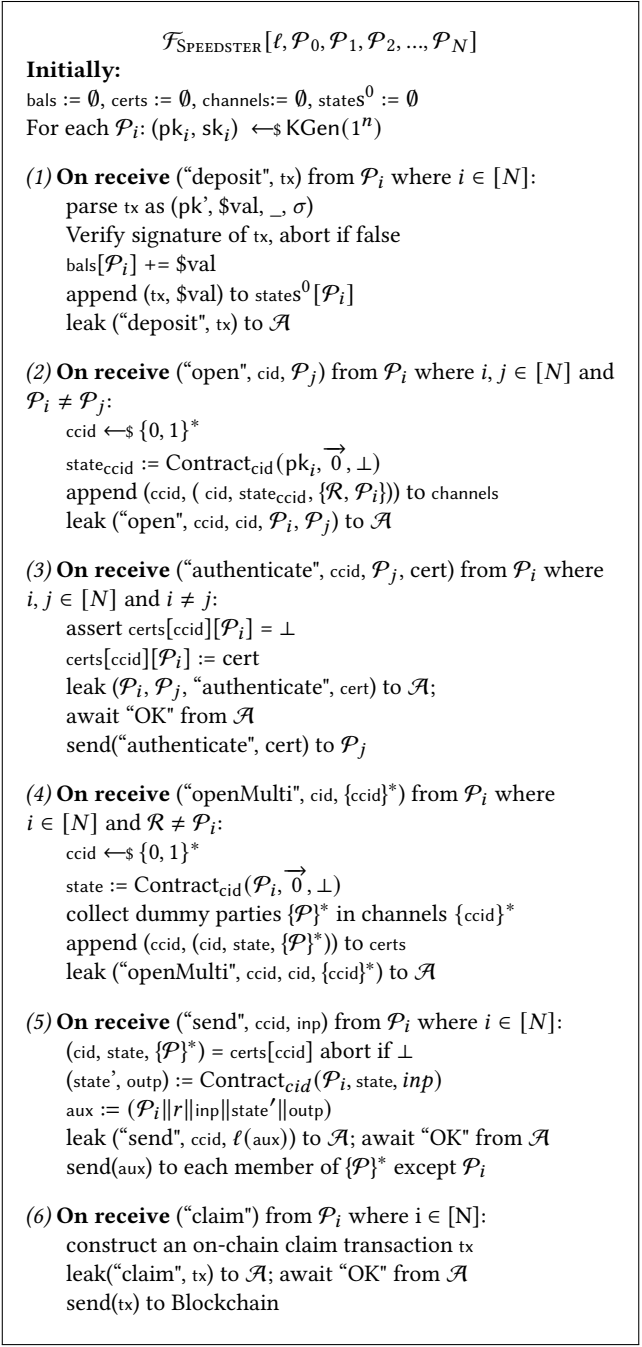

\begin{framed}
 \begin{flushleft}
\begin{center}
  
$\mathcal{F}_{\TheName{}}[\ell,\idealP{}_{\mathcolor{0}}, \idealP{}_{\mathcolor{1}}, \idealP{}_{\mathcolor{2}}, ... ,\idealP{}_{\mathcolor{N}}]$
        
\end{center}

\noindent \textbf{Initially:} 

\sffont{bals} := $\emptyset$,
\sffont{certs} := $\emptyset$,
\sffont{channels}:= $\emptyset$,
\spstate s$^0$ := $\emptyset$

For each \idealPi: ($\pk_i$, $\sk_i$) $\sample \kgen(\secparam)$

\bigskip
\setlength\parindent{15pt}
\noindent \colorgray{(1)} \textbf{On receive} (``deposit", \tx) from \idealPi{} where $i \in [N]$:

parse \tx{} as (\pk', \$val, \_, $\sigma$)

Verify signature of \tx{}, abort if false

\sffont{bals}[\idealPi] += \$val

append (\tx, \$val) to \spstate s$^0[\idealPi]$

leak (``deposit", \tx) to \idealA{}

\bigskip

\noindent \colorgray{(2)} \textbf{On receive} (``open", \cid, \idealPj) from \idealPi{} where $i, j \in [N]$ and $ \idealPi{} \neq \idealPj$: 

\ccid $\sample \bin^*$

\spstate$_{\ccid}$ := Contract$_{\cid}(\pk_i, \overrightarrow{0}, \perp)$

append (\ccid, ( \cid, $\spstate_{\ccid}$, \{\idealR, \idealPi \})) to $\sffont{channels}$

leak (``open", \ccid, \cid, \idealPi, \idealPj) to \idealA{}

\bigskip

\noindent \colorgray{(3)} \textbf{On receive} (``authenticate", \ccid, \idealPj, cert) from \idealPi{} where $i, j \in [N]$ and $i \neq j$: 

assert \sffont{certs}[\ccid][\idealPi] = $\perp$

\sffont{certs}[\ccid][\idealPi] := cert

leak (\idealPi, \idealPj, ``authenticate", \cert) to $\idealA$;

await ``OK" from \idealA{}

send(``authenticate", cert) to \idealPj{}

\bigskip

\noindent \colorgray{(4)} \textbf{On receive} (``openMulti", \cid, \{\ccid\}$^*$) from \idealPi{} where $i \in [N]$ and \idealR{} $\neq \idealPi$: 

\ccid $\sample \bin^*$

\spstate{} := Contract$_{\cid}(\idealPi, \overrightarrow{0}, \perp)$

collect dummy parties \{\idealP{}\}$^*$ in channels $\{\ccid\}^*$

append (\ccid, (\cid, \spstate, \{\idealP{}\}$^*$)) to $\sffont{certs}$

leak (``openMulti", \ccid, \cid, \{\ccid\}$^*$) to \idealA{}

\bigskip

\noindent \colorgray{(5)} \textbf{On receive} (``send", \ccid, \inp) from \idealPi{} where $i \in [N]$:

(\cid, \spstate, \{\idealP{}\}$^*$) = \sffont{certs}[\ccid] abort if $\perp$

(\spstate', \outp) := Contract$_{cid}(\idealPi, \spstate, inp)$

$\msg := (\idealPi\concat r\concat \inp\concat \spstate'\concat \outp)$

leak (``send", \ccid, $\ell(\msg)$) to \idealA{}; await ``OK" from \idealA{}



send(\msg) to each member of \{\idealP\}$^*$ except \idealPi{}






\bigskip

\noindent \colorgray{(6)} \textbf{On receive} (``claim") from \idealPi{} where i $\in$ [N]:

construct an on-chain claim transaction \tx{}

leak(``claim", \tx) to \idealA; await ``OK" from \idealA{}

send(\tx) to Blockchain

\end{flushleft}
\end{framed}
\caption{Ideal functionality of $\mathcal{F}_{\TheName{}}$. Internal communications are assumed to be encrypted with authenticated encryption.}
\label{fig:Ideal_Main}
\end{figure}

While dealing with a corrupted party \idealPi{}:
\begin{itemize}
    
    \item \idealS{} queries a set of channel id \{\ccid\}$^*$ and a smart contract id from \idealE{}. Then, \idealS{} sends the message (``openMulti", \cid{}, \{\ccid\}$^*$) to \idealMain{} and collects the output with \idealPi{}'s identity. Then \idealS{} emulates a ``resume" call to \idealGatt{} with the same messages on behalf of \idealPi{} and collects the output from \idealGatt{}.
    \item Upon receiving message (``openMulti", \ccid{}, \cid{}, \{\ccid{}\}$^*$) from \idealMain{}. \idealS{} emulates a ``resume" call to \idealGatt{} sending message (``openMulti", \cid{}, \{\ccid{}\}$^*$). Then relay the output to \idealPi{}.
    
\end{itemize}

\vspace{3pt}
\noindent (5) \textit{Channel Transaction:} Upon receiving message (``send", \ccid{}, $\ell(\msg)$) from \idealMain{} of \idealPi{}, \idealS{} requests a $\sffont{key}$ from a challenger \sffont{Ch} who generates \ASE{} keys. \idealS{} generates a random string $r$, and computes $m := \mathcal{AE}.\enc(\sffont{key}, r)$, of which $\abs{m} = \abs{\ell(\msg)}$. \idealS{} emulates a ``resume" call to \idealGatt{} sending message (``receive", $\ccid{}, m$) on behalf of \idealPi{}. Then relay the output to \idealPi{}.

While dealing with a corrupted party \idealPi{}:
\begin{itemize}
    
    \item \idealS{} queries a channel id \ccid{} and a random string \inp{} := \bin$^*$ from \idealE{}. Then, \idealS{} sends the message (``send", \cid{}, \{\ccid{}\}$^*$) to \idealMain{} on \idealPi{}'s behalf, and collects the output. Then \idealS{} emulates a ``resume" call to \idealGatt{} with the same messages on behalf of \idealPi{} and collects the output from \idealGatt{}.
    
    \item Upon receiving message (``send", \ccid{}, $\ell(\msg)$) from \idealMain{}. \idealS{} requests a \sffont{key} from $\sffont{Ch}$. \idealS{} computes $m := \mathcal{AE}.\enc(\sffont{key}, \overrightarrow{0})$, of which $\abs{m} = \abs{\ell(\msg)}$. \idealS{} emulates a ``resume" call to \idealGatt{} sending message (``receive", \ccid{}, $m$) on behalf of \idealPi{}. Then relay the output to \idealPi{}.
    
\end{itemize}

\vspace{3pt}
\noindent (6) \textit{Claim:} Upon receiving message (``claim", \tx{}) of \idealPi{} from \idealMain{}, \idealS{} emulates a ``resume" call to \idealGatt{} sending message (``claim", \tx{}) on behalf of \idealPi{}. Then, and send ``OK" to \idealMain{}, and relay the output to the Blockchain.

While \idealPi{} is corrupted. \idealS{} sends message (``claim") to \idealMain{} on behalf of \idealPi{} and collects the output. Then \idealS{} emulates a ``resume" call to \idealGatt{} with the same message on behalf of \idealPi{} and collects the output from \idealGatt{}, then relay the output to the Blockchain.

\begin{figure}[htbp]
\begin{framed}
\begin{flushleft}   
   \begin{center}
    $\mathcal{G}_{att}[\Sigma, reg]$
    \end{center}
    
\texttt{// initialization:}

\setlength\parindent{5pt}

\noindent \textbf{On initialize}: $(\mpk, \msk) := \Sigma.\kgen(\secparam); T = \emptyset$

\bigskip

\noindent \texttt{// public query interface:}

\noindent \textbf{On receive}\text{*} getpk() from some \idealP{}: send \mpk{} to \idealP{}

\begin{center}
\noindent\line(1,0){220}
\end{center}

\begin{center}
\textbf{Enclave operations}
\end{center}

\noindent \texttt{//local interface — install an enclave:}

\noindent \textbf{On receive}\text{*} install($idx, \sffont{prog}$) from some $\idealP{} \in reg$:

if \idealP{} is honest, assert $idx = sid$

generate nonce $eid \in \bin^\lambda$, 

store $T[eid, \idealP{}] := (idx, \sffont{prog}, ~0)$, send $eid$ to \idealP{}

\bigskip

\noindent \texttt{// local interface — resume an enclave:}

\noindent \textbf{On receive}\text{*} resume($eid, inp, switch:=on$) from some $\idealP{} \in reg$:

let $(idx, \sffont{prog}, mem) := T[eid, \idealP{}]$, abort if not found

let $(outp, mem) := \sffont{prog}(inp, mem),$

update $T[eid, \idealP{}] := (idx , \sffont{prog}, mem)$

if $switch$ is set to $on$
\setlength\parindent{15pt}

let $\sigma := \Sigma.\sig_{\msk{}}(idx, eid, \sffont{prog}, outp)$

send $(outp, \sigma)$ to \idealP{}

\setlength\parindent{5pt}
otherwise:
\setlength\parindent{15pt}
 
send $(outp, \perp)$ to \idealP{}

\end{flushleft}
\end{framed}
    \caption{A global functionality modeling an SGX-like secure processor. Compared to \cite{pass2017formal}, a switch is added to the ``resume" command to allow users to disable the signature. The default value of $switch$ is set to ``on".}
    \label{fig:gatt}
\end{figure}
\begin{figure}[htbp]
\begin{framed}
\begin{flushleft}   
   \begin{center}
    $\mathcal{F}_{blockchain}[Contract]$
    \end{center}
    
\texttt{// initialization:}

\setlength\parindent{5pt}

\noindent \textbf{On initialize}: $Storage := \emptyset$

\bigskip

\noindent \texttt{// public query interface:}

\noindent \textbf{On receive}\text{*} read($id$) from $\idealP{}$:

output $Storage[id]$, or $\perp$ if not found

\bigskip

\noindent \texttt{// public append interface:}

\noindent \textbf{On receive}\text{*} append(\tx) from $\idealP{}$:

abort if $Storage[\tx.id]$ $\neq \perp$

if $Contract(\tx) = true:$

\setlength\parindent{15pt}
$Storage[\tx.id] := \tx$

output ("success")

\setlength\parindent{5pt}
else

\setlength\parindent{15pt}
output ("failure")

\end{flushleft}
\end{framed}
    \caption{Ideal functionality of Blockchain modeling an append-only ledger. }
    \label{fig:blockchain}
\end{figure}

\end{appendix}

\end{document}